\documentclass[longauth,final]{aa}
\usepackage[english]{babel}

\usepackage{amsmath}
\usepackage{multirow}
\usepackage[colorlinks=true, allcolors=blue]{hyperref}
\usepackage{natbib}

\newcommand{\fer}{{\sl {\it Fermi}}}
\newcommand{\fla}{\fer-LAT}

\usepackage{lineno}
\renewcommand{\thelinenumber}{}

\begin{document}
\nolinenumbers
\title{Detection of the distant quasar OP 313 with the first Large-Sized Telescope of CTAO}
\titlerunning{Detection of the FSRQ OP 313 at VHE}

\author{
\footnotesize
K.~Abe\inst{1} \and
S.~Abe\inst{2} \and
J.~Abhir\inst{3} \and
A.~Abhishek\inst{4} \and
V.~A.~Acciari\inst{5} \and
F.~Acero\inst{6,7} \and
A.~Aguasca-Cabot\inst{8} \and
I.~Agudo\inst{9} \and
C.~Alispach\inst{10} \and
D.~Ambrosino\inst{11,12}
% ...
\and
(the MAGIC and CTAO-LST Collaborations) 
\and
I. Myserlis\inst{107,108} \and
R. Rao\inst{109} \and
M. Gurwell\inst{109} \and
G. Keating\inst{109}\and
A. Marscher\inst{110}\and
S. Jorstad\inst{110}\and
E. Angelakis\inst{111}\and
A. Kraus\inst{108}\and
C. Thum\inst{108}\and
J. A. Acosta-Pulido\inst{17}\and
A. Marchini\inst{112}\and
L. Stiaccini\inst{4}\and
P. Aceti\inst{113}\and
M. Banfi\inst{113}\and
S. Leonini\inst{114}\and
M. Conti\inst{114}\and
P. Rosi\inst{114}\and
L. M. Tinjaca Ramirez\inst{114}\and
J. Escudero Pedrosa\inst{109,9}\and
A. Sota\inst{9}\and
V. Casanova\inst{9}\and
F. J. Aceituno\inst{9}\and
V. Fallah Ramazani\inst{97}
}

\institute{
Department of Physics, Tokai University, 4-1-1, Kita-Kaname, Hiratsuka, Kanagawa 259-1292, Japan
\and
Institute for Cosmic Ray Research, University of Tokyo, 5-1-5, Kashiwa-no-ha, Kashiwa, Chiba 277-8582, Japan
\and
ETH Zürich, CH-8093 Zürich, Switzerland
\and
INFN and Universit\`a degli Studi di Siena, Dipartimento di Scienze Fisiche, della Terra e dell'Ambiente (DSFTA), Sezione di Fisica, Via Roma 56, 53100 Siena, Italy
\and
Institut de Fisica d'Altes Energies (IFAE), The Barcelona Institute of Science and Technology, Campus UAB, 08193 Bellaterra (Barcelona), Spain
\and
Universit\'e Paris-Saclay, Universit\'e Paris Cit\'e, CEA, CNRS, AIM, F-91191 Gif-sur-Yvette Cedex, France
\and
FSLAC IRL 2009, CNRS/IAC, La Laguna, Tenerife, Spain
\and
Departament de Física Quàntica i Astrofísica, Institut de Ciències del Cosmos, Universitat de Barcelona, IEEC-UB, Martí i Franquès, 1, 08028, Barcelona, Spain
\and
Instituto de Astrofísica de Andalucía-CSIC, Glorieta de la Astronomía s/n, 18008, Granada, Spain
\and
Department of Astronomy, University of Geneva, Chemin d'Ecogia 16, CH-1290 Versoix, Switzerland
\and
INFN Sezione di Napoli, Via Cintia, ed. G, 80126 Napoli, Italy
\and
INAF - Osservatorio Astronomico di Roma, Via di Frascati 33, 00040, Monteporzio Catone, Italy
\and
Università di Udine and INFN Trieste, I-33100 Udine, Italy
\and
also at International Center for Relativistic Astrophysics (ICRA), Rome, Italy
\and
Max-Planck-Institut für Physik, Boltzmannstraße 8, 85748 Garching bei München, Germany
\and
INFN Sezione di Padova and Universit\`a degli Studi di Padova, Via Marzolo 8, 35131 Padova, Italy
\and
Instituto de Astrof\'isica de Canarias and Departamento de Astrof\'isica, Universidad de La Laguna, C. V\'ia L\'actea, s/n, 38205 La Laguna, Santa Cruz de Tenerife, Spain
\and
Univ. Savoie Mont Blanc, CNRS, Laboratoire d'Annecy de Physique des Particules - IN2P3, 74000 Annecy, France
\and
Croatian MAGIC Group: University of Zagreb, Faculty of Electrical Engineering and Computing (FER), 10000 Zagreb, Croatia
\and
Saha Institute of Nuclear Physics, A CI of Homi Bhabha National Institute, Kolkata 700064, West Bengal, India
\and
Universit\"at Hamburg, Institut f\"ur Experimentalphysik, Luruper Chaussee 149, 22761 Hamburg, Germany
\and
Graduate School of Science, University of Tokyo, 7-3-1 Hongo, Bunkyo-ku, Tokyo 113-0033, Japan
\and
IPARCOS-UCM, Instituto de F\'isica de Part\'iculas y del Cosmos, and EMFTEL Department, Universidad Complutense de Madrid, Plaza de Ciencias, 1. Ciudad Universitaria, 28040 Madrid, Spain
\and
Faculty of Science and Technology, Universidad del Azuay, Cuenca, Ecuador
\and
Institut f\"ur Theoretische Physik, Lehrstuhl IV: Plasma-Astroteilchenphysik, Ruhr-Universit\"at Bochum, Universit"atsstra\ss e 150, 44801 Bochum, Germany
\and
Centro Brasileiro de Pesquisas F\'isicas, Rua Xavier Sigaud 150, RJ 22290-180, Rio de Janeiro, Brazil
\and
Faculty of Physics and Applied Informatics, University of Lodz, ul. Pomorska 149-153, 90-236 Lodz, Poland
\and
CIEMAT, Avda. Complutense 40, 28040 Madrid, Spain
\and
INFN Sezione di Bari and Politecnico di Bari, via Orabona 4, 70124 Bari, Italy
\and
INAF - Osservatorio Astronomico di Brera, Via Brera 28, 20121 Milano, Italy
\and
Departament de F\'isica Qu\`antica i Astrof\'isica, Institut de Ci\`encies del Cosmos, Universitat de Barcelona, IEEC-UB, Mart\'i i Franqu\`es, 1, 08028, Barcelona, Spain
\and
Aix Marseille Univ, CNRS/IN2P3, CPPM, Marseille, France
\and
National Institute for Astrophysics (INAF), I-00136 Rome, Italy
\and
INAF - Osservatorio di Astrofisica e Scienza dello spazio di Bologna, Via Piero Gobetti 93/3, 40129 Bologna, Italy
\and
INFN Sezione di Trieste and Universit\`a degli studi di Udine, via delle scienze 206, 33100 Udine, Italy
\and
University of Geneva - D\'epartement de physique nucl\'eaire et corpusculaire, 24 Quai Ernest Ansernet, 1211 Gen`eve 4, Switzerland
\and
Departament de Física, and CERES-IEEC, Universitat Autònoma de Barcelona, E-08193 Bellaterra, Spain
\and
INAF - Istituto di Astrofisica e Planetologia Spaziali (IAPS), Via del Fosso del Cavaliere 100, 00133 Roma, Italy
\and
Università di Pisa and INFN Pisa, I-56126 Pisa, Italy
\and
INFN Sezione di Bari and Universit\`a di Bari, via Orabona 4, 70126 Bari, Italy
\and
INFN Sezione di Torino, Via P. Giuria 1, 10125 Torino, Italy
\and
Dipartimento di Fisica - Universitá degli Studi di Torino, Via Pietro Giuria 1 - 10125 Torino, Italy
\and
Palacky University Olomouc, Faculty of Science, 17. listopadu 1192/12, 771 46 Olomouc, Czech Republic
\and
Dipartimento di Fisica e Chimica 'E. Segr\`e' Universit\`a degli Studi di Palermo, via delle Scienze, 90128 Palermo, Italy
\and
INFN Sezione di Catania, Via S. Sofia 64, 95123 Catania, Italy
\and
also at Como Lake centre for AstroPhysics (CLAP), DiSAT, Universit\`a dell'Insubria, via Valleggio 11, 22100 Como, Italy
\and
Department for Physics and Technology, University of Bergen, Norway
\and
IRFU, CEA, Universit'e Paris-Saclay, B\^at 141, 91191 Gif-sur-Yvette, France
\and
Instituto de Física de Sao Carlos, Universidade de Sao Paulo, Av. Trabalhador Sao-carlense, 400 - CEP 13566-590, Sao Carlos, SP, Brazil
\and
University of Alcal\'a UAH, Departamento de Physics and Mathematics, Pza. San Diego, 28801, Alcal'a de Henares, Madrid, Spain
\and
also at Port d'Informaci\'o Cient\'ifica (PIC), E-08193 Bellaterra (Barcelona), Spain
\and
INFN Sezione di Bari, via Orabona 4, 70125, Bari, Italy
\and
University of Rijeka, Department of Physics, Radmile Matejcic 2, 51000 Rijeka, Croatia
\and
Institute for Theoretical Physics and Astrophysics, Universit\"at W\"urzburg, Campus Hubland Nord, Emil-Fischer-Str. 31, 97074 W\"urzburg, Germany
\and
Department of Physics, TU Dortmund University, Otto-Hahn-Str. 4, 44227 Dortmund, Germany
\and
INFN Sezione di Roma La Sapienza, P.le Aldo Moro, 2 - 00185 Rome, Italy
\and
ILANCE, CNRS – University of Tokyo International Research Laboratory, University of Tokyo, 5-1-5 Kashiwa-no-Ha Kashiwa City, Chiba 277-8582, Japan
\and
Technische Universität Dortmund, D-44221 Dortmund, Germany
\and
Physics Program, Graduate School of Advanced Science and Engineering, Hiroshima University, 1-3-1 Kagamiyama, Higashi-Hiroshima City, Hiroshima, 739-8526, Japan
\and
INFN Sezione di Roma Tor Vergata, Via della Ricerca Scientifica 1, 00133 Rome, Italy
\and
Armenian MAGIC Group: ICRANet-Armenia, 0019 Yerevan, Armenia
\and
University of Split, FESB, R. Bo\v{s}kovi'ca 32, 21000 Split, Croatia
\and
Department of Physics, Yamagata University, 1-4-12 Kojirakawa-machi, Yamagata-shi, 990-8560, Japan
\and
Physics Program, Graduate School of Advanced Science and Engineering, Hiroshima University, 1-3-1 Kagamiyama, Higashi-Hiroshima City, Hiroshima, 739-8526, Japan
\and
Sendai College, National Institute of Technology, 4-16-1 Ayashi-Chuo, Aoba-ku, Sendai city, Miyagi 989-3128, Japan
\and
Université Paris Cité, CNRS, Astroparticule et Cosmologie, F-75013 Paris, France
\and
Josip Juraj Strossmayer University of Osijek, Department of Physics, Trg Ljudevita Gaja 6, 31000 Osijek, Croatia
\and
INFN Dipartimento di Scienze Fisiche e Chimiche - Universit`a degli Studi dell'Aquila and Gran Sasso Science Institute, Via Vetoio 1, Viale Crispi 7, 67100 L'Aquila, Italy
\and
Chiba University, 1-33, Yayoicho, Inage-ku, Chiba-shi, Chiba, 263-8522 Japan
\and
Kitashirakawa Oiwakecho, Sakyo Ward, Kyoto, 606-8502, Japan
\and
Finnish MAGIC Group: Finnish Centre for Astronomy with ESO, Department of Physics and Astronomy, University of Turku, FI-20014 Turku, Finland
\and
FZU - Institute of Physics of the Czech Academy of Sciences, Na Slovance 1999/2, 182 21 Praha 8, Czech Republic
\and
Laboratory for High Energy Physics, \'Ecole Polytechnique F\'ed\'erale, CH-1015 Lausanne, Switzerland
\and
Astronomical Institute of the Czech Academy of Sciences, Bocni II 1401 - 14100 Prague, Czech Republic
\and
Faculty of Science, Ibaraki University, 2 Chome-1-1 Bunkyo, Mito, Ibaraki 310-0056, Japan
\and
Sorbonne Université, CNRS/IN2P3, Laboratoire de Physique Nucléaire et de Hautes Energies, LPNHE, 4 place Jussieu, 75005 Paris, France
\and
Graduate School of Science and Engineering, Saitama University, 255 Simo-Ohkubo, Sakura-ku, Saitama city, Saitama 338-8570, Japan
\and
also at Department of Physics, University of Oslo, Norway
\and
Institute of Particle and Nuclear Studies, KEK (High Energy Accelerator Research Organization), 1-1 Oho, Tsukuba, 305-0801, Japan
\and
University of Geneva, Chemin d’Ecogia 16, CH-1290 Versoix, Switzerland
\and
INFN Sezione di Trieste and Universit\`a degli Studi di Trieste, Via Valerio 2 I, 34127 Trieste, Italy
\and
Croatian MAGIC Group: University of Rijeka, Faculty of Physics, 51000 Rijeka, Croatia
\and
Escuela Polit'ecnica Superior de Ja\'en, Universidad de Ja\'en, Campus Las Lagunillas s/n, Edif. A3, 23071 Ja\'en, Spain
\and
Institute for Nuclear Research and Nuclear Energy, Bulgarian Academy of Sciences, 72 boul. Tsarigradsko chaussee, 1784 Sofia, Bulgaria
\and
Inst. for Nucl. Research and Nucl. Energy, Bulgarian Academy of Sciences, BG-1784 Sofia, Bulgaria
\and
Dipartimento di Fisica e Astronomia (DIFA) Augusto Righi, Università di Bologna, via Gobetti 93/2, I-40129 Bologna, Italy
\and
Department of Physics and Astronomy, Clemson University, Kinard Lab of Physics, Clemson, SC 29634, USA
\and
Grupo de Electronica, Universidad Complutense de Madrid, Av. Complutense s/n, 28040 Madrid, Spain
\and
E.S.CC. Experimentales y Tecnología (Departamento de Biología y Geología, Física y Química Inorgánica) - Universidad Rey Juan Carlos
\and
Macroarea di Scienze MMFFNN, Università di Roma Tor Vergata, Via della Ricerca Scientifica 1, 00133 Rome, Italy
\and
Institute of Space Sciences (ICE, CSIC), and Institut d'Estudis Espacials de Catalunya (IEEC), and Instituci'o Catalana de Recerca I Estudis Avan\c{c}ats (ICREA), Campus UAB, Carrer de Can Magrans, s/n 08193 Bellatera, Spain
\and
Department of Physics, Konan University, 8-9-1 Okamoto, Higashinada-ku Kobe 658-8501, Japan
\and
School of Allied Health Sciences, Kitasato University, Sagamihara, Kanagawa 228-8555, Japan
\and
RIKEN, Institute of Physical and Chemical Research, 2-1 Hirosawa, Wako, Saitama, 351-0198, Japan
\and
INFN MAGIC Group: INFN Sezione di Torino and Università degli Studi di Torino, I-10125 Torino, Italy
\and
Finnish MAGIC Group: Space Physics and Astronomy Research Unit, University of Oulu, FI-90014 Oulu, Finland
\and
Charles University, Institute of Particle and Nuclear Physics, V Hole\v{s}ovi\v{c}k'ach 2, 180 00 Prague 8, Czech Republic
\and
Division of Physics and Astronomy, Graduate School of Science, Kyoto University, Sakyo-ku, Kyoto, 606-8502, Japan
\and
Institute for Space-Earth Environmental Research, Nagoya University, Chikusa-ku, Nagoya 464-8601, Japan
\and
Kobayashi-Maskawa Institute (KMI) for the Origin of Particles and the Universe, Nagoya University, Chikusa-ku, Nagoya 464-8602, Japan
\and
Graduate School of Technology, Industrial and Social Sciences, Tokushima University, 2-1 Minamijosanjima,Tokushima, 770-8506, Japan
\and
INFN Sezione di Pisa, Edificio C – Polo Fibonacci, Largo Bruno Pontecorvo 3, 56127 Pisa, Italy
\and
INAF Istituto di Astrofisica Spaziale e Fisica Cosmica di Palermo, Via Ugo La Malfa 153, Palermo, I-90146, Italy
\and
also at INAF Padova
\and
Gifu University, Faculty of Engineering, 1-1 Yanagido, Gifu 501-1193, Japan
\and
Department of Physical Sciences, Aoyama Gakuin University, Fuchinobe, Sagamihara, Kanagawa, 252-5258, Japan
\and 
Institut de Radioastronomie Millimétrique, Avenida Divina Pastora, 7, Local 20, E–18012 Granada, Spain
\and Max-Planck-Institut für Radioastronomie, Auf dem Hügel 69, D-53121 Bonn, Germany
\and Center for Astrophysics | Harvard \& Smithsonian, 60 Garden Street, Cambridge, MA 02138, USA
\and Institute for Astrophysical Research, Boston University, 725 Commonwealth Ave., Boston, MA 02215, USA
\and Orchideenweg 8, 53123 Bonn, Germany
\and Astronomical Observatory, Department of Physical Sciences, Earth and Environment, University of Siena, Siena, I-53100, Italy
\and Osservatorio Astronomico Città di Seveso, Seveso, I-20822, Italy
\and Montarrenti Observatory, Siena, I-53100, Italy
}
\authorrunning{CTAO-LST Project, MAGIC Collaboration and MWL collaborators}

  \abstract
   {Flat Spectrum Radio Quasars (FSRQs) often remain undetected at very-high-energy (VHE, $E > 100$~GeV) $\gamma$ rays due to their typically soft spectra and large distance. In December 2023, the Large-Sized Telescope prototype (LST-1) detected for the first time VHE $\gamma$-ray emission from the FSRQ OP~313 during an extraordinarily bright flare, becoming the furthest blazar ever observed at VHE with $z=0.997$.}
   {We aim to characterize the $\gamma$-ray emission of OP 313 during this flare, comparing it with its average emission state in order to understand the processes leading to this detection. Its remarkable distance also enables studies on the Extragalactic Background Light (EBL), with the goal of evaluating the attenuation of VHE $\gamma$-ray photons. Thanks to an intensive multi-wavelength campaign, we also study in detail the broadband emission of the source.}
   {We characterize the $\gamma$-ray emission during the flare in December 2023 and the low emission state observed in January 2024 thanks to the LST-1 and MAGIC data and quasi-simultaneous \textit{Fermi}-LAT observations.{This dataset also enables us to evaluate the EBL attenuation by systematically exploring the EBL intensity over the $\gamma$-ray spectrum.}. Finally, we study the multi-wavelength emission and interpret the broadband spectral energy distribution (SED) within blazar radiative models including the thermal contributions from the accretion disc, dusty torus and broad line region. }
   {We measure significant VHE $\gamma$-ray emission during the December 2023 flare at a level of 0.3 Crab Units {above 100 GeV}. We also characterize the flare brightness in the high-energy (HE, $E>100$~MeV) $\gamma$-ray band, that was found to be a factor 50 above the average emission seen by \textit{Fermi}-LAT. The HE and VHE observations allow us to set constrains to the EBL density, resulting in a upper limit at 0.6~$\mu$m of $\lambda I_{\lambda}<8.74$~nW m$^{2}$ sr$^{-1}$ with LST-1 data, and $\lambda I_{\lambda}<14.7$~nW m$^{2}$ sr$^{-1}$ with MAGIC data, including systematic uncertainties. Finally, thanks to the extensive multi-wavelength campaign organized, we are able to construct and model the broadband SED of OP~313 within the framework of a two-zone leptonic model where the $\gamma$-ray emission {is produced via inverse-Compton scattering of the broad line region, accretion disk and dusty torus photon fields. However the dominant external photon field remains unknown, as several combinations are able to successfully explain the $\gamma$-ray emission observed}.}
   {}

\offprints{lst-contact@cta-observatory.org. $^{\star}$Corresponding authors (in alphabetical order): A.~Arbet-Engels, J.~Baxter, L.~Heckmann, D.~Morcuende, M.~Nievas~Rosillo, J.~Otero-Santos, D.~A.~Sanchez.} 

\keywords{Gamma rays: galaxies -- Galaxies: active -- Galaxies: jets -- Galaxies: nuclei -- quasars: individual: OP 313 -- radiation mechanisms: non-thermal}

\maketitle

\section{Introduction}\label{sec1}
Active Galactic Nuclei (AGNs) are {some} of the most luminous, violent and variable objects in the Universe, consisting of a supermassive black hole (SMBH) with a typical mass $>$10$^6 M_{\odot}$ that is {persistently} accreting matter from its surroundings. 
A fraction of these AGNs develop powerful relativistic jets characterized by strong non-thermal continuum emission that can extend from radio frequencies to very-high-energy (VHE, $E>100$~GeV) $\gamma$ rays. 
These `radio-loud' AGNs are the dominant class of extragalactic VHE $\gamma$-ray emitters, with about 100 sources having being detected as of now\footnote{see \url{http://tevcat.uchicago.edu} for a detailed and updated list of known VHE emitting sources.}. 
The vast majority of these VHE AGNs have relativistic jets closely aligned with the line of sight, collectively referred to as blazars, yet they are an heterogeneous group with different fundamental and observational properties.

Blazars are often sub-classified in BL Lacertae objects (BL Lacs), with {radiatively} inefficient accretion flows and dominated by continuum non-thermal emission from accelerated particles within the jet, and Flat Spectrum Radio Quasars (FSRQs), more luminous as a group, with richer AGN structures, and correspondingly more complex radiation fields \citep{urry1995}. Apart from the SMBH and the relativistic jet, the canonical picture of {an} FSRQ includes an efficient accretion flow (accretion disc, AD), sometimes visible in the optical-UV band as a ``big blue bump'' \citep[see][]{jolley2009}; fast moving gas clouds from the so called broad line region (BLR) producing strong broad emission lines arising from the reprocess of the disk emission; and a dusty torus (DT) with strong absorption of the optical light coming from the central regions of the AGN and its corresponding re-emission as an infrared thermal component.

The broadband multi-wavelength Spectral Energy Distribution (SED) of blazars typically shows a characteristic double bump shape that is interpreted with a combination of different physical emission mechanisms. The low-energy emission --- extending from radio to UV-X-rays --- is well-explained as synchrotron radiation of relativistic electrons moving under the influence of the magnetic field of the relativistic jet \citep{koenigl1981}. On the other hand, the origin of the high-energy emission is more debated, with two scenarios being considered. The most used framework is based on leptonic processes via Inverse Compton (IC) scattering of low-energy photons with relativistic electrons. This IC can occur either with the relativistic electrons interacting with low-energy synchrotron photons, i.e. synchrotron self-Compton \citep[SSC, see][]{maraschi1992}, or with low-energy seed photons from outside the jet through the so-called external Compton scattering \citep[EC, see][]{dermer1993}. However, and especially owing to the need of hadronic processes to explain rare neutrino emission events, hadronic models have also been developed, where processes such as proton synchrotron or Bethe-Heitler pair production are invoked to explain the high-energy spectral component {\citep[e.g.][]{2024arXiv241114218C}.}

OP~313 is a FSRQ located at redshift $z=0.997$ \citep{schneider2010}. Past observations by the Major Atmospheric Gamma-ray Imaging Cherenkov (MAGIC) telescopes on this target resulted only in upper limits \cite[see][]{magic2024}. After the report {of} the high emission state observed in optical \citep{otero2023} and high-energy (HE, $E>100$~MeV) $\gamma$ rays \citep{bartolini2023}, and based on daily analysis of the \fla\ data using the FLaapLUC pipeline \citep{2018A&C....22....9L}, the object was observed by the Large-Sized Telescope prototype (LST-1) starting on December 10\textsuperscript{th} 2023, and resulting in the detection of the source in the VHE band \citep{cortina2023}.
OP~313 is the tenth FSRQ ever detected at VHE $\gamma$ rays as well as the most distant AGN in this energy range to date, surpassing  PKS~0346-27 \citep[z=0.991,][]{pks_0346}, B0218+357 \citep[z=0.954,][]{b0218_magic}, and PKS~1441+25 \citep[z=0.939,][]{pks1441_magic,pks1441_veritas}, all of them also FSRQs. In addition to LST-1, observations from MAGIC were also triggered following the high state detected by \textit{Fermi}-LAT, and started on December 10\textsuperscript{th}. The MAGIC analysis confirmed the VHE detection of LST-1. We remark that in May 2012, the IceCube observatory measured a neutrino track-like event consistent with the position of OP~313 \citep{aartsen2016}, drawing a special attention to this blazar. The event was not temporally coincident with a $\gamma$-ray flare \citep{Marinelli2021}. Additional multi-wavelength studies are key to decipher the potential multimessenger nature of OP~313.
 
In this work, we characterize the broadband emission of OP 313 between December 2023 and January 2024, where the source transitioned from a very bright optical and $\gamma$-ray state into an intermediate state. 
We discuss the implications of the detection of significant VHE $\gamma$-ray emission from this source to constrain the intensity of the Extragalactic Background Light (EBL). We also present an extensive multi-wavelength campaign organized during the observations performed by LST-1, that allows us to evaluate the behaviour of the source across all the spectrum. Finally, and supported by this intense multi-wavelength observing campaign, we provide possible interpretations of the broadband emission of OP~313 within leptonic blazar theoretical emission scenarios. 

\section{Observations}\label{sec2} 
\subsection{LST-1}\label{sec2.1}
LST-1 is the first of the four LSTs that will be part of the Cherenkov Telescope Array Observatory (CTAO) in the Northern Hemisphere site \citep{acharyya2019} at the Roque de los Muchachos Observatory in the Canary island of La Palma (Spain). LST-1 is equipped with a 23-m diameter mirror dish and a camera composed of 1855 high-quantum-efficiency photomultipliers. These characteristics provide LST-1 with a large collection area and high sensitivity, enabling the detection of $\gamma$-ray events down to energies of $\sim$20~GeV \citep{abe2023}, and making LST-1 the ideal instrument for the observation of distant VHE $\gamma$-ray emitters. Its best integral sensitivity is at a level of 1.1\% of the Crab Nebula flux above 250 GeV after 50 h of observations and, on average, it is approximately 1.5 times less sensitive than the {two MAGIC telescopes} above 100 GeV \citep{abe2023}.

\begin{figure*}[ht!]
\centering
\includegraphics[width=\textwidth]{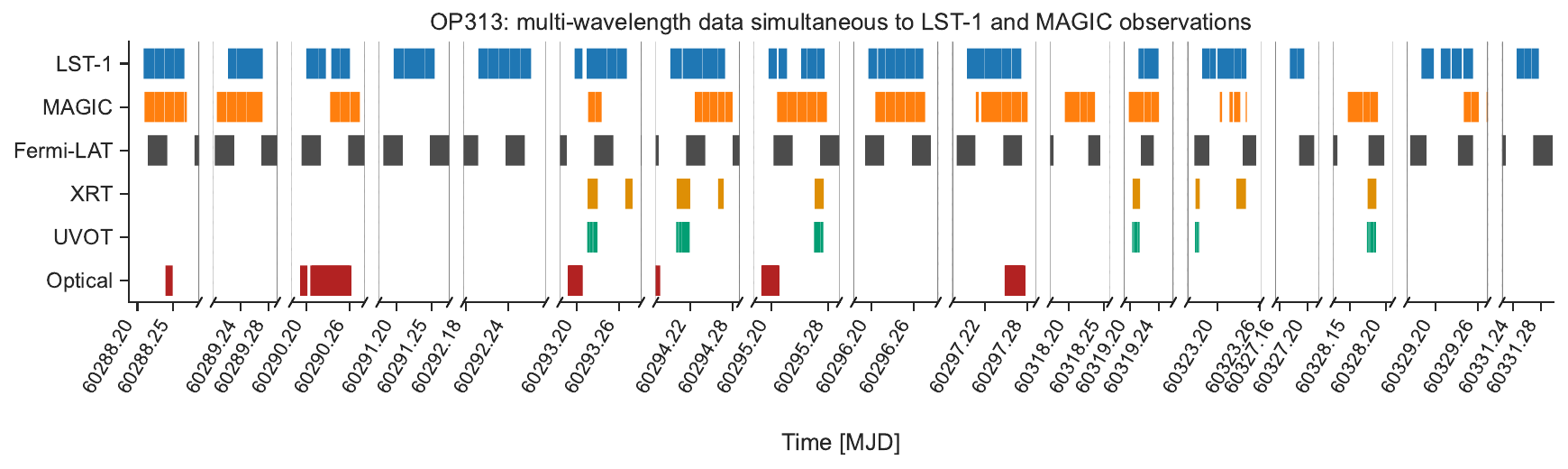}
\caption{Good time intervals (GTIs) of the OP 313 observations performed in December 2023 and January 2024 from VHE $\gamma$-ray instruments (LST-1 and MAGIC) and strictly simultaneous data from \textit{Fermi}-LAT, \textit{Swift}-XRT, \textit{Swift}-UVOT, and optical facilities.}
\label{fig:gtis}
\end{figure*}

LST-1 observed OP~313 in the so-called wobble pointing mode with a wobble offset angle of 0.4 deg \citep{wobble_mode_fomin} in December 2023 following high-energy $\gamma$-ray \citep{bartolini2023} and optical \citep{otero2023} alerts, leading to the detection of the source after four nights of observations \citep{cortina2023}. Following the initial detection, LST-1 continued observing OP~313, accumulating a total of $\simeq$15 hours of data after quality cuts between December 10\textsuperscript{th} and December 19\textsuperscript{th}. Moreover, about 5 additional hours of data were collected during January 2024, when the source was found to be in a lower emission state, resulting in no significant VHE $\gamma$-ray emission. All data were taken in dark conditions in zenith angles ranging from $\sim$10$^{\circ}$ to $\sim$55$^{\circ}$. The periods of good quality data (good time intervals, GTIs) taken by LST-1 are shown in Fig.~\ref{fig:gtis}.

\subsection{MAGIC}\label{sec2.2}
{The Florian Goebel MAGIC telescopes\footnote{\url{https://magic.mpp.mpg.de/}}} form a system of two 17-m diameter Imaging Atmospheric Cherenkov Telescopes (IACTs) located at $\sim$2200\,m above sea level on the Canary island of La Palma within the Roque de los Muchachos Observatory. The stereoscopic system is sensitive to an energy range between tens of GeVs to tens of TeVs with a sensitivity of about 2\% (about 1\%) of the Crab Nebula flux above 100~GeV (300 GeV) at low zenith angles ($<$~30$^\circ$) after 25~h of observations \citep[see Fig.~19 of][]{2016APh....72...76A}. After the high state in $\gamma$-ray reported by \cite{bartolini2023} during November 2023, MAGIC observed OP 313 between December 10\textsuperscript{th} 2023 to December 19\textsuperscript{th} 2023 (MJD~60288 to MJD~60297) for a total of $\sim$10\,hours after data quality cuts based on atmospheric transmission. Additional observations were performed in 2024 between January 9th to 20th (MJD~60318 to MJD~60329) resulting in $\sim$4\,hours after data quality cuts. These observations were performed in wobble mode with a 0.4-deg offset angle.

\subsection{Fermi-LAT}\label{sec2.3}
The Large Area Telescope \citep[LAT,][]{atwood2009} on-board the \fer\ satellite is a pair-conversion telescope designed to detect high-energy (HE) $\gamma$-ray photons from 50~MeV to 1 TeV \citep{atwood2009}. Launched in 2008, the LAT typically operates in survey mode, providing full-sky coverage every 3 hours. 

We collected \textit{Fermi}-LAT \texttt{Pass 8 SOURCE} events with energies larger than $100\,$MeV within a region of interest (ROI) of $15\,$deg radius around the OP~313 position from December 1, 2023, to January 23, 2024. The sky model has been generated based on the fourth \fla\ catalogue of sources \citep[4FGL, ][]{4fgl,4fgl_dr3} and a binned analysis has been performed.

\subsection{Swift-XRT}\label{sec2.4}
The XRT X-ray telescope \citep{burrows2004} is the main instrument onboard the Neil Gehrels Swift Observatory ({\it Swift}). It is sensitive to photons of energies in the range of $0.3-10\,$~keV.

Since OP~313 is not a very bright source in the X-ray band, count rates of $\leq 0.2\,$c/s are typical and therefore no pile-up is expected. 
Consequently, most of the data from OP~313 from XRT is taken in photon-counting mode to retain full imaging and spectroscopy. 
{\it Swift}-XRT data {from} OP~313 were downloaded from \textit{Swift} database\footnote{\url{https://www.swift.ac.uk/swift_portal/}}.
Of all available data from the telescope, this work focuses on data collected in December 2023 and January 2024 with Swift observation IDs 00036384039 to 00036384061. 

\subsection{Swift-UVOT}\label{sec2.5}
The \textit{Swift}-UVOT \citep{roming2005} is an optical-UV telescope onboard {\it Swift}. The detector
is equipped with two grisms for low resolution spectroscopy and six primary filters (V, B, U, UVW1, UVM2, and UVW2), which cover the wavelength range from $1600$ to $6000\,$\AA. For most of the observation ids in Sect.~\ref{sec2.4}, OP~313 was observed with the full six-filter set of UVOT. Similar to XRT data, we downloaded UVOT data from the \textit{Swift} database.
Moreover, some of these observations are affected by a non-optimal pointing and tracking accuracy as a result of the degradation of one IRO (gyroscope) as described in \cite{gcn1,gcn2}. 

\subsection{Optical observations}\label{sec2.6}
An extensive optical campaign was organized in order to provide coverage at optical wavelengths to the VHE $\gamma$-ray observations, involving a number of optical telescopes and observatories around the globe. The Sierra Nevada Observatory (SNO) in Granada (Spain) conducted a photo-polarimetric follow-up of the source during December 2023 and January 2024, thanks to its 1.5-m (T150) and 0.9-m (T90) telescopes. The T150 telescope is equipped with the AndorT150 instrument, with the capability of performing $BVRI$ optical photometry. On the other hand, the T90 relied on the DIPOL-1 polarimeter \citep{otero2024} to follow the polarized emission of the source during this period in the optical $R$ band. Both sets of observations were performed under the TOP-MAPCAT blazar photo-polarimetric monitoring programme \citep{agudo2012}, relying on the automatic photo-polarimetric analysis \textsc{python} pipeline IOP4 \citep{escudero2023,escudero2024}. 

Observations with the 0.8-m IAC80 telescope at Teide Observatory (Tenerife, Spain). The IAC80, equipped with the CAMELOT2\footnote{\url{https://research.iac.es/OOCC/iac-managed-telescopes/iac80/camelot2-2/}} camera, provided data in the optical $BRI$ bands. Optical observations were also scheduled and coordinated with the robotic network of telescopes from Las Cumbres Observatory\footnote{\url{https://lco.global}} \citep{brown2013}. These observations were conducted with the 0.4-m class telescopes and the SCICAM QHY600 instrument \citep{harbeck2024}, using the $r$-band filter.

Optical data from the Tuorla blazar monitoring programme\footnote{\url{https://users.utu.fi/kani/1m/index.html}} conducted by the University of Turku (Finland) in the $R$ band, currently with the 0.8-m Joan Oró Telescope at Montsec Astronomical Observatory (Spain), were also included in the dataset \citep{nilsson2018}. Moreover, data in the $VRI$ bands were taken and provided by the 0.3-m telescope at Siena, and by the Seveso and Montarrenti Observatories during the high emission state in December 2023. Additional $R$-band data were obtained from the observations of the 1.5-m Kanata Telescope at the Higashi-Hiroshima Observatory, Hiroshima Astrophysical Science Center, Hiroshima University. Observations were primarily conducted with the HONIR instrument \citep{2012SPIE.8446E..73S,2014SPIE.9147E..4OA} mounted on the Kanata Telescope, which allows simultaneous observations in both the $R$ band and near-IR band, with data in the near-IR $J$ band also incorporated into the overall observational dataset.

Data from several optical public surveys and databases that observed OP 313 during this period were also retrieved. The All-Sky Automated Survey for Supernovae\footnote{\url{https://asas-sn.osu.edu}} \citep[ASAS-SN,][]{shappee2014,kochanek2017}, located in Hawaii and focused on supernovae observations, provided optical data in the Sloan $g$ band owing to its regular sky monitoring roughly once every five nights. The Zwicky Transient Facility \citep[ZTF,][]{bellm2019,masci2019}, aimed for a systematic study of the optical sky approximately every two nights from Palomar Observatory (California, US), observed OP 313 in the optical Sloan $gri$ bands, that were retrieved from its publicly available database\footnote{\url{https://www.ztf.caltech.edu}}. Finally, data from the Asteroid Terrestrial-impact Last Alert System (ATLAS) project\footnote{\url{https://atlas.fallingstar.com}} \citep{tonry2018} were also included in this work. In particular, ATLAS, focused on asteroid detection, but again containing data on sources of every type due to their all-sky scanning nature with telescopes located in Hawaii, Chile and South Africa, provided data for OP~313 in the optical band with the $c$ (4200-6500~\AA) and $o$ (5600-8200~\AA) wide bandpass filters.

\subsection{Radio and millimetric observations}\label{sec2.7}
Observations in several radio-millimetric bands were coordinated during the period considered in this study with the Submillimeter Array \citep[SMA,][]{ho2004} radio telescope at Mauna Kea (Hawaii) and the 100-m Effelsberg radio telescope located in Germany. SMA is an 8-element 6-meter diameter dish ratio interferometer. It has two orthogonally polarized receivers, enabling also polarimetric measurements of the radio flux. These receivers are inherently linearly polarized, however they can be converted to circular thanks to the quarter-wave plates installed on the polarimeter \citep{marrone2008}. SMA observations were performed as part of the SMA Monitoring of AGNs with Polarization (SMAPOL) programme \citep{myserlis2025}, dedicated to a regular monitoring of 40 $\gamma$-ray bright blazars at millimetric wavelengths. A total of eight observations have been performed by SMA between December 2023 and January 2024 with integration times between 3 and 27 minutes. Six of the observations correspond to the flaring state observed during December --- two of them within the LST-1 and MAGIC observing windows --- and the two {remaining observation} performed during the low $\gamma$-ray emission period in January 2024, only one of them being performed before the last observation of LST-1 and MAGIC in the considered period.

The 100-m Effelsberg radio telescope has also contributed to follow the radio emission of the source within the framework of the Monitoring the Stokes $Q$, $U$, $I$ and $V$ Emission of AGN jets in Radio (QUIVER) programme \citep{krauss2003,myserlis2018,myserlis2025}, working from 2.6 GHz to 44 GHz (11 to 7 mm) and following the radio emission of $\gamma$-ray blazars including OP~313. The Effelsberg radio telescope is equipped with six receivers with two orthogonally polarized feeds that can provide, apart from total flux estimates in the radio band, polarimetric observables of the emission of the observed target. OP~313 was observed twice in the 4.85-GHz, 10.45-GHz and 14.25-GHz bands during the December 2023 outburst, and two more times during the low VHE $\gamma$-ray emission state from January 2024. As for the SMA data, only one of the two observations corresponding to the low state where performed within the campaign considered here.

\section{Data analysis}\label{sec3}

\subsection{LST-1}\label{sec3.1}

The analysis of the LST-1 raw data was done with the \texttt{cta-lstchain} software \citep{lstchain_adass,lopez-coto2024} {based on \texttt{ctapipe} \citep{ctapipe},} where calibration, data quality selection, data cleaning, parametrization, $\gamma$-hadron separation, direction, energy reconstruction and event selection are performed, following the standard source-independent analysis described in \cite{abe2023}. The event classification uses Random Forest trained with Monte Carlo (MC) simulated proton and $\gamma$-ray events along a declination of 34.76$^{\circ}$, close to that of OP~313. The Instrument Response Functions (IRFs) of the telescope were estimated by interpolating the IRFs calculated at different sky positions distributed along all the sky, where MC events are simulated. Energy-dependent cuts were optimized to preserve 70\% of the $\gamma$-ray-like events both on {the probability to be a $\gamma$-ray event and angular distance of the reconstructed direction to the nominal location of OP~313 (gammaness and $\theta$).} 

\begin{figure}[ht!]
\centering
\includegraphics[width=\columnwidth]{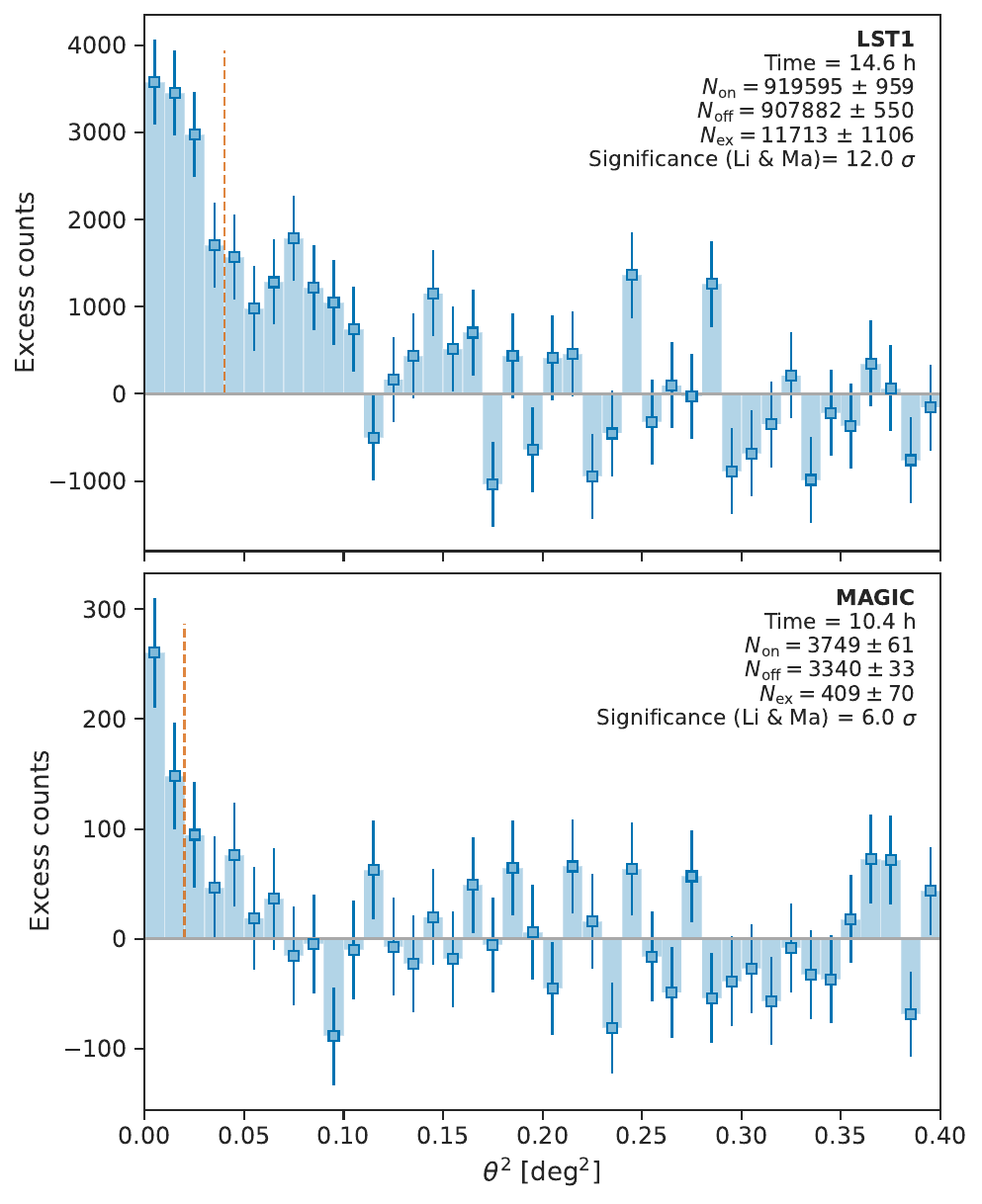}
\caption{Distributions of the squared angular distance of the reconstructed $\gamma$-ray directions to the nominal location of OP 313 during December 2023, derived from the excess $\gamma$-ray events. \textit{Top:} LST-1. \textit{Bottom:} MAGIC. In the case of LST-1, we obtained the distribution for energies {above 10\,GeV}. The dashed line delimits the region where events are considered to calculate the statistical significance of the detection. These signal regions are defined as the standard $\theta^2$ cuts for each instrument, which are related to their PSF, as detailed in \cite{abe2023} for LST-1 and \cite{2016APh....72...76A} for MAGIC.}
\label{fig:theta2}
\end{figure}

The high-level data analysis of the selected $\gamma$-ray events was performed with the \textsc{python} package \texttt{Gammapy} \citep{donath2023,gammapy_version}. In order to characterize and subtract the sky background, 3 {OFF regions located at the same offset from the FoV center} were used. The energy threshold of $\sim$60~GeV was calculated for this analysis by weighting the default MC spectrum by a spectral index 4, preliminarily obtained for this source without considering the EBL absorption.
A safe mask on the effective area $A_{\rm eff}$ of 5\% of $A_{\rm eff}^{\rm max}$ was also considered for estimating this energy threshold, and applied to each individual observation run.
The $\theta^{2}$ distribution used to assess the detection significance is shown in Fig.~\ref{fig:theta2} ({top} panel). OP~313 exhibits a clear excess during the high state of December 2023, with a statistical significance of $12.0\sigma$ {above $\sim10\,\mathrm{GeV}$}. {This excess corresponds to a 1.3\% signal-over-background level. The level of background systematics and the corresponding impact on the reconstructed flux are evaluated in Appendix~\ref{sec:systematics}.} This highlights the intrinsic challenge of detecting sources near threshold energies with a single Cherenkov telescope, where the observations are fully background-limited. In contrast, no significant excess was detected during January 2024, when the source was in a low flux state.

\begin{figure*}
\centering
\includegraphics[width=0.9\linewidth]{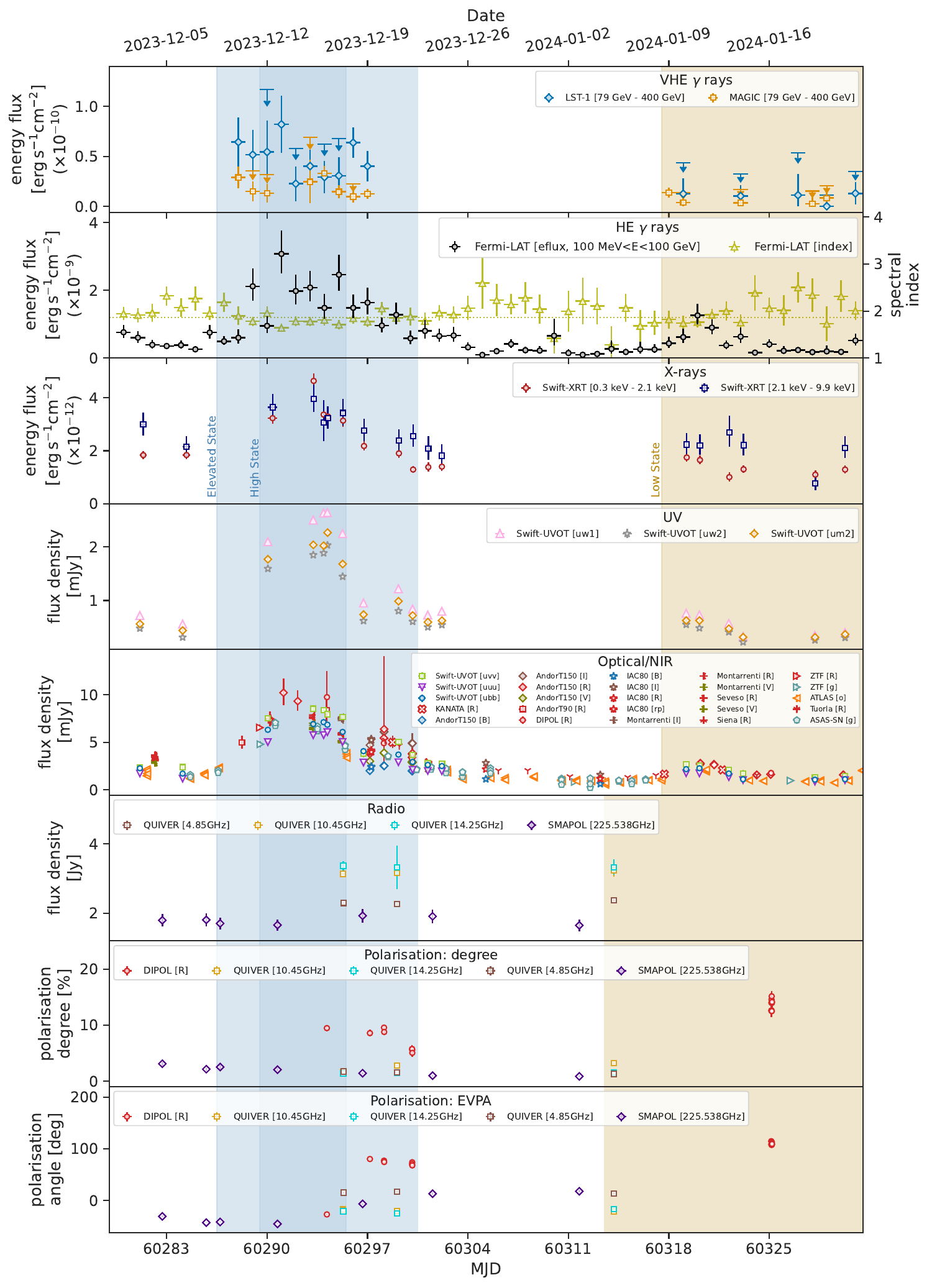}
\caption{\label{fig:MWL_LCs}Multi-wavelength lightcurves of OP 313. \textit{From top to bottom:} VHE $\gamma$ rays above 79 GeV (LST-1 and MAGIC); HE $\gamma$ rays between 100 MeV and 300 GeV (\textit{Fermi}-LAT); X-rays (\textit{Swift}-XRT); UV (\textit{Swift}-UVOT); optical and IR bands (\textit{Swift}-UVOT, ZTF, ASAS-SN, SNO, IAC80, Las Cumbres, Tuorla, Kanata, Montarrenti, Siena, Seveso, ZTF and ATLAS); radio-millimetric (QUIVER and SMAPOL); polarization degree (SNO, QUIVER and SMAPOL); and polarization angle (SNO, QUIVER and SMAPOL). Light blue band: extended high state (December 2023) used for EBL estimation. Dark blue band: multi-wavelength stable high state used for broadband SED modeling. Light brown band: low state (January 2024).
}
\end{figure*}

The spectrum and Spectral Energy Distribution (SED) of OP~313 were characterized by using a compound spectral shape comprising a power-law function and a term accounting for the strong EBL absorption due to its large distance, using the EBL model from \cite{saldana-lopez2021}.
The best fit to this spectral shape was later used to calculate the integral flux above 100~GeV and the light curve shown in Fig.~\ref{fig:MWL_LCs}. 
The best fit model is a simple power-law. 
{More complex intrinsic spectral models like a log-parabola or a power law with exponential cut-off do not result in significantly better fit quality, hence there is no evidence for intrinsic spectral curvature in the energy range probed by LST-1.}
The results of the LST-1 spectrum for the high state are reported in Table~\ref{table:VHE_analysis} and the spectrum of December 2023 is displayed in Fig.~\ref{fig:JointSED}.

The LST-1 VHE light curve is presented in the top panel of Fig.~\ref{fig:MWL_LCs}, where the temporal evolution of the integrated flux above an energy of 100~GeV is shown. Statistics were maximized via pre-stacking same-night runs together before running the light curve estimation code. To reduce the effect of the source's spectral shape on the flux calculation, we fixed a power-law index of 2.7 for both LST-1 and MAGIC in the light curve estimation---a value very close to the mid-point between the spectral indices reconstructed by the two instruments. 
We show night-wise flux points for all measurements over a significance of 2$\sigma$. For those observations below that significance value, we show also the 95\% confidence level upper limits.
The average emission above 100 GeV during the high state period of December 2023 is at a level of 0.3 Crab Units (C.U.)\footnote{We refer along the paper to the Crab Nebula flux as measured by the MAGIC telescopes \citep{2016APh....72...76A}.}, equivalent to $\sim 1.5 \times 10^{-10}$~cm$^{-2}$~s$^{-1}$.

Systematic effects affecting the LST-1 data analysis were evaluated, focusing on those related to the background estimation, absolute energy scale and event selection cuts. Each of these effects constitute a contribution to the systematic uncertainties of 30\%, (+57\%, -47\%) and 23\% of the flux above 100~GeV, respectively.  
Systematics on the energy scale yield an uncertainty on the spectral index of $-0.31, +0.44$. A detailed discussion on the estimation of these uncertainties is reported in Appendix~\ref{sec:systematics}. 

\begin{table*}[htp]
\caption{Results of the VHE LST-1 and MAGIC analyses} 
\label{table:VHE_analysis}
\begin{center}
\resizebox{\textwidth}{!}{
\begin{tabular}{c c c c c c c c c}
\hline\hline 
\multirow{2}{*}{Period} & \multirow{2}{*}{Telescope} & Obs. & $E_{th}$ & Li \& Ma  & $\phi_{\rm 0}(E_{\rm 0})$  & \multirow{2}{*}{$\Gamma$} & $F_{\gamma}$ ($>100$~GeV) & $p$-value  \\

 & & time [h] & [GeV]  & signif. &  [$10^{-8}\,{\rm cm^{-2}}\,{\rm s^{-1}}\,{\rm TeV^{-1}}$]    &  & [$10^{-10}\,{\rm cm^{-2}}\,{\rm s^{-1}}$] & (from MC)  \\\hline
\hline

\multirow{2}{*}{December} & LST-1 & 14.6 & $\sim 60$ & 12.0 & $1.15 \pm 0.14$ & $1.99 \pm 0.55$ & $1.50 \pm 0.21$ & $\sim 0.57$\\
                          & MAGIC & 10.4 & $\sim 70$ & 6.0  & $0.39 \pm 0.07$ & $3.33 \pm 0.64$ & $0.41 \pm 0.09$ & $\sim 0.88$ \\\hline
\end{tabular}
}
\end{center}
\tablefoot{{Power-law spectral shape and EBL absorption} with the model from \cite{saldana-lopez2021} as been assumed. Reference energy is 100 GeV. The spectral index and reference flux correspond to the deabsorbed, intrinsic spectrum. {The reported errors correspond to statistical uncertainties only. A detailed evaluation of systematic effects is reported in Appendix~\ref{appA.1}.}}
\end{table*}

\subsection{MAGIC}\label{sec3.2}
MAGIC data were processed using the MARS software package \citep[MAGIC Analysis and Reconstruction Software;][]{zanin2013, 2016APh....72...76A} following the standard data analysis procedure. Figure~\ref{fig:theta2} (bottom panel) shows the resulting $\theta^2$ plot for the December 2023 observations. We select events {using a gammaness cut efficiency of 72\%}, and the detection significance is calculated by defining a signal region of $\theta^2<0.02$. These are the standard cuts in the MAGIC analyses focused on the low energies. We further apply a cut on the \textit{size} parameter (number of photoelectrons in the event image) which is tuned to optimize the signal at the lowest energies. Indeed, owing to the large redshift \citep[$z=0.997$; ][]{schneider2010} resulting in a soft spectrum caused by the EBL absorption above $100$\,GeV, most of the source events are expected near the low-energy threshold. To optimize the \textit{size} cut, we use MC events rescaled to match a power-law distribution of index $4.5$, which is the (observed) slope of the source measured with MAGIC prior to any cut optimization. We then scanned several \textit{size} cuts and selected the value that provides the highest significance using Eq. (17) of \citet{li1983}. While the $\gamma$-ray signal events are taken from MCs, the background events are taken from the observations themselves. A \textit{size} cut of 30 provides the highest significance, and this cut is applied for the rest of the MAGIC analysis in this work. At a zenith angle of $25^\circ$, being the lowest zenith angle at which OP~313 was observed, we obtain an analysis threshold of $\approx$70\,GeV.

As shown in the bottom panel of Fig.~\ref{fig:theta2}, the source is detected by MAGIC with a significance of 6.0 $\sigma$. The signal-over-background level is more than $10\%$, meaning the MAGIC measurement is limited by flux level and sensitivity, rather than by background systematic uncertainties as in the case of LST-1. Regarding January 2024 observations, the source is not significantly detected, suggesting a lower activity. The significance of the signals are calculated using Eq. (17) of \citet{li1983}.

To compute the spectra, we applied energy dependent gammaness and $\theta^2$ cuts with an efficiency of 90\% and 75\%, respectively. We modelled the MAGIC data in \texttt{Gammapy} using a power-law function and correcting for the EBL absorption using the model of \cite{saldana-lopez2021}. 
{As in the LST-1 analysis, we applied to each individual run a safe mask on the effective area $A_{\rm eff}$. For MAGIC, the threshold was lowered to 1\% of $A_{\rm eff}^{\rm max}$ to avoid excessive event suppression at low energies, where the stereoscopic reconstruction leads to a steeper effective area curve. This prevents the removal of the lowest significant energy bin, which would otherwise make the analysis unfeasible for such a steep-spectrum, distant source. This choice comes at the cost of additional systematic uncertainties, which are discussed in \ref{appA.2}.}
The MAGIC spectrum does not show a significant preference for a log-parabola model with respect to a power-law model. The resulting spectral parameters for the December period during  which a significant signal is obtained are shown in Table~\ref{table:VHE_analysis}. The MAGIC lightcurve was computed above 0.1\,TeV using this best-fit model, following the same strategy to maximize statistics and reduce systematics as in LST-1. The resulting light curve is shown in Fig.~\ref{fig:MWL_LCs}. For nights with a significance smaller than 2$\sigma$ we computed ULs at a 95\% confidence level.

We evaluated uncertainties related to the systematics on the absolute energy scale as well as on the background estimation. The details of those investigations are reported in Appendix~\ref{sec:systematics}. In summary, the systematics on the energy scale yield an uncertainty on the $>100$\,GeV flux of +74\%, -52\%, while the systematics related to the background normalization is at the level of {20-30}\%. 
Systematics on the energy scale yield an uncertainty on the spectral index of $-0.32, +0.16$.

\subsection{Fermi-LAT}\label{sec3.3}
Data binned on a daily basis (centred at midnight UTC), have been analysed and stored into {\tt Gammapy} datasets \citep{nievas2025}. This allows a flexible analysis by only stacking the daily-bins matching LST-1 or MAGIC observations, ensuring contemporaneous coverage.

Both a power-law and log-parabola spectral models where tested for OP~313. Using the likelihood ratio test, we found the best model for each data set. Whenever the log-parabola model is not preferred at more that 2$\sigma$ with respect to the power-law, the later model was used.

The results of the analysis, obtained using daily bins, are presented in Table~\ref{table:Fermi} together with the corresponding bins. Restricting to data taken $\pm$1 hour around the observing window of each IACTs yields reconstructed fluxes compatible with the ones presented in this work, although with larger statistical uncertainties. 

\begin{table*}[htp]
\caption{\textit{Fermi}-LAT only analysis, matching the observation nights of LST-1 and MAGIC respectively.} 
\centering
\begin{tabular}{c c c c c c}
\hline\hline 
\multirow{2}{*}{Name} &\multirow{2}{*}{MJD} &$\phi_{\rm 0}(E_{\rm 0})$  & \multirow{2}{*}{$\alpha$}& \multirow{2}{*}{$\beta$}  \\
 &   & [$10^{-8}\,{\rm cm^{-2}}\,{\rm s^{-1}}\,{\rm GeV^{-1}}$]   & &    \\\hline
\hline 

\multirow{4}{*}{LST-1 December} & 60287.5, 60288.5, 60289.5 & \multirow{4}{*}{4.06 $\pm$  0.31} & \multirow{4}{*}{1.70 $\pm$  0.08} & \multirow{4}{*}{0.17 $\pm$  0.05}  \\ 
&  60290.5, 60291.5, 60292.5 &&& \\
&  60293.5, 60294.5, 60295.5&&& \\
&  60296.5 &&& \\
\multirow{3}{*}{MAGIC December} &  60287.5, 60288.5, 60289.5 & \multirow{3}{*}{3.70 $\pm$  0.32} & \multirow{3}{*}{1.77 $\pm$  0.09} & \multirow{3}{*}{0.14 $\pm$  0.05}  \\
& 60292.5, 60293.5, 60294.5 &&& \\
&  60295.5, 60296.5&&& \\\hline

 \multirow{2}{*}{LST-1 January} &60318.5, 60322.5, 60326.5, & \multirow{2}{*}{1.22 $\pm$  0.20} & \multirow{2}{*}{2.07 $\pm$  0.14} & \multirow{2}{*}{0.20 $\pm$  0.02}  \\
 & 60328.5, 60330.5 &&& \\
 \multirow{2}{*}{MAGIC January} & 60317.5, 60318.5, 60322.5& \multirow{2}{*}{1.28   $\pm$  0.20} &  \multirow{2}{*}{2.14  $\pm$ 0.16} &\multirow{2}{*}{0.34 $\pm$ 0.09}  \\
 & 60327.5, 60328.5 &&& \\\hline

\end{tabular}
\tablefoot{The columns give the list of days considered in the analysis (in lower edge MJD notation), the differential flux at a reference energy of 2~GeV, the spectral index $\alpha$ and the curvature parameters $\beta$.}
\label{table:Fermi}
\end{table*}

\subsection{Swift-XRT}\label{sec3.4}
The data were processed using the standard Heasoft's {\tt xrtproducts} task \citep{2009MNRAS.397.1177E} to extract spectra from calibrated and screened event files. 
The signal and background extraction regions defined as circular and annuli around the coordinates of OP 313, using radii of $r_{\rm on}=40$ and $r_{\rm off,in}=80$, $r_{\rm off,out}=300$ arcseconds, respectively. 
The final product is an OGIP-compliant spectral dataset for each observation ID, which was subsequently analysed using \texttt{Gammapy}. We correct the Galactic hydrogen absorption considering a hydrogen column density of $N_H = 1.25 \times 10^{20}$~cm$^{-2}$ \citep{2000ApJ...542..914W}.

\subsection{Swift-UVOT}\label{sec3.5}
Due to the high brightness of OP~313 in these bands, the detection of the source was possible in the 6 bands of UVOT, and a standard aperture photometry method could be applied to extract the source flux. Signal and background regions were defined, with a circular aperture of $25\,$ arcseconds and annular background region with radii of $40$ and $100$ arcseconds respectively, both larger than the recommended sizes due to the mentioned degradation in the pointing and tracking accuracy. The data were processed using the {\tt uvot2pha} task to generate the final spectral files in OGIP format, one per filter and observation ID, which later are ingested into \texttt{Gammapy} datasets. 
After data reduction, all data have been corrected by Galactic extinction using the extinction values derived from \cite{cardelli1989} for each band with a colour excess $E_{B-V}=0.0144$ \citep{1998ApJ...500..525S}.

\subsection{Optical data}\label{sec3.6}
The optical data were reduced and analyzed following standard differential aperture photometric procedures. The signal of the target is evaluated by defining an aperture in which its signal is computed, and an annular region around it to determine the background. The sizes of the aperture and annular regions vary between different observatories depending on the specific configuration of each analysis. Reference stars of known, constant magnitude and colour relations in the {field of view} are then used for calibrating the magnitude of the source.
More details on the standard steps of the optical photo-polarimetric analysis can be consulted for instance in \cite{nilsson2018} and references therein. We checked and corrected possible calibration offsets between data taken in each filter by different telescopes by comparing strictly simultaneous observations. As done for the UVOT data, all observations have been corrected by Galactic extinction with the model from \cite{cardelli1989}. Finally, and owing to the very large redshift of OP~313, we note that no subtraction of the host galaxy emission has been performed as it is very weak compared to the continuum synchrotron emission.

Regarding the polarization data, we applied the common correction to the electric vector position angle (EVPA) to {account} for the $\pm 180^{\circ}$ degeneracy in the definition of the angle \citep[e.g.][]{kiehlmann2016,otero-santos2023}. For this, we add/subtract $n \times 180^{\circ}$ to an EVPA measurement $i$ to minimize the difference with the previous point $i-1$ whenever $| \theta_i - \theta_{i-1} | > 90^{\circ}$. For this correction, despite restricting our dataset to the period mentioned above, we take into account older data from the photo-polarimetric TOP-MAPCAT programme so the correction is applied in consistency with observations prior to December 2023.

\subsection{Radio-millimetric data}\label{sec3.7}

The SMA and Effelsberg data were reduced and calibrated following standard procedures \citep[see for instance][]{terasranta1998}. SMAPOL data reduction was performed using the \texttt{MIR} software\footnote{\url{https://lweb.cfa.harvard.edu/~cqi/mircook.html}}. MWC 349 A, Callisto, Uranus, Neptune and Ceres were used as total flux calibrators. For the analysis of the polarized emission, instrumental polarization leakage was calibrated for the lower and upper sidebands using the \texttt{gpcal} task \citep{sault1995}, and removed from the data. On the other hand, the Effelsberg data were calibrated as detailed by \cite{myserlis2018}. The instrumental polarization of QUIVER observations was characterized and calibrated using observations of polarized and unpolarized calibrators and removed from the data as specified in \cite{myserlis2018}. Then, the polarization intensity, degree and angle for the data from both telescopes were obtained from the Stokes I, Q and U parameters.

\section{Results}\label{sec4}

\subsection{$\gamma$-ray SED and extragalactic background light constraints}\label{sec:sed_ebl}

The EBL \citep{Dwek_2013} is the cumulative background radiation across the infrared, optical, and UV regions of the electromagnetic spectrum, representing the total light from all extragalactic sources over the history of the Universe. Direct measurements of the EBL are challenging due to atmospheric effects and strong local foregrounds (e.g., zodiacal light, diffuse galactic emission, dust extinction), so it is often inferred through modeling \citep[e.g.,][]{dominguez_2023new,saldana-lopez2021} or indirect methods, such as the approach presented here using $\gamma$-ray data.

\begin{figure}
\centering 
\includegraphics[width=\columnwidth]{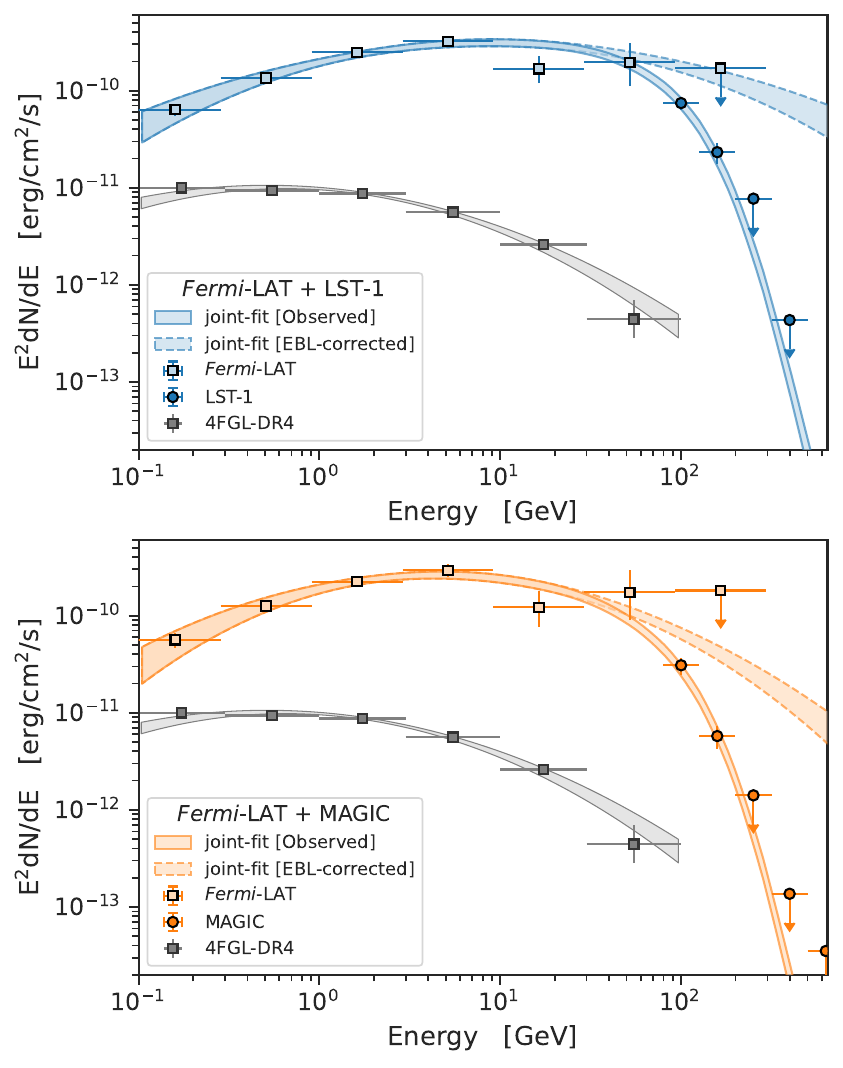}
\caption{\label{fig:JointSED}Spectral energy distribution of OP 313 in the $\gamma$-ray range. \textit{Top}: LST-1 data taken in December and contemporaneous \fla\ data. \textit{Bottom}: MAGIC data taken in December and contemporaneous \fla\ data. 1$\sigma$ intervals are marked as colored bands with solid lines, while the intrinsic emission (EBL-corrected) is marked with lighter bands within dashed contour lines. Flux points are also represented, with squared markers and filled circles for {\it Fermi}-LAT and LST-1 or MAGIC respectively  Finally, the quiescent state of the source is represented through its flux points (grey square markers) and best-fit model (grey band) from the 4FGL-DR4 catalog \citep{4fgl_dr3,4fgl_dr4}.}
\end{figure}

The attenuation of very-high-energy (VHE) $\gamma$ rays by EBL photons is modeled as $\text{F}_{\text{obs}} = \exp(-\tau(z, E_\gamma)) \times \text{F}_{\text{int}},$ where $\text{F}_{\text{obs}}$ is the observed flux, $\text{F}_{\text{int}}$ is the intrinsic flux, and $\tau(z, E_\gamma)$ is the optical depth for the propagation of $\gamma$ rays. Following previous works \citep{2010ApJ...723.1082A,2012Sci...338.1190A,2013A&A...550A...4H,2017A&A...606A..59H,2019MNRAS.486.4233A}, we introduce a dimensionless scaling factor $\alpha$ such that the absorption term becomes $\exp(-\alpha \tau(z, E_\gamma))$. In principle, $\alpha$ is independent of both the energy $E_\gamma$ and the redshift $z$. 

To evaluate $\alpha$, we fit the observational spectra (binned event data) using the baseline EBL model from \citet{saldana-lopez2021} combined with smooth, simple spectral functions to model the intrinsic $\gamma$-ray emission, such as a log-parabola (LP), a power law with an exponential cutoff (PLEC), and a log-parabola with an exponential cutoff (LPEC). The assumption of these spectral forms adequately reproducing the intrinsic source spectrum is critical because it underpins the application of Wilks’ theorem \citep{Wilks:1938dza} to our likelihood ratio tests. In other words, if these spectral models faithfully represent the intrinsic emission, then the change in the -2 log-likelihood is asymptotically $\chi^2$-distributed, which justifies the derivation of confidence intervals on the EBL parameters.

To validate this point, we first examined the observed spectrum of OP 313 over the energy range from $100\,$MeV to $1\,$TeV by jointly fitting the ``LST-1 December'' \textit{Fermi}-LAT and LST-1 datasets (see Table \ref{table:Fermi}) as well as the corresponding ``MAGIC December'' \textit{Fermi}-LAT and MAGIC datasets. In these fits, we combined the proposed simple spectral shapes with the nominal EBL attenuation model from \citet{saldana-lopez2021}. By selecting specific nights for stacking the \textit{Fermi}-LAT nights that tightly match the VHE observations, we ensured that the source was in a comparable state of activity across instruments, thus minimizing systematic biases due to source variability. A simple visual inspection of Fig. \ref{fig:JointSED} suggests that a log-parabolic spectral model combined with EBL attenuation adequately describes the observed data. In fact, the derived flux points for \textit{Fermi}-LAT, LST-1, and MAGIC closely follow the best-fit model, with minimal residuals and no indication that a more complex spectral form is needed. As an additional verification, we tested the LPEC model, which has one degree of freedom more than the LP model, and {compatible} results were obtained. The corresponding model parameters for the \textit{Fermi}-LAT + LST-1 and \textit{Fermi}-LAT + MAGIC LP fits are summarized in Table \ref{tab:sedfit}.

We then systematically fit the observed spectra by fixing $\alpha$ to values between 0 and 3 in steps. For each fixed $\alpha$, we perform the joint analysis, storing the best-fit model and the corresponding likelihood (see Figure \ref{fig:EBL_Likelihood_profiles}). By analyzing the relative difference in the 2$\times$log-likelihood values, we derived upper limits on $\alpha$, and hence on the EBL intensity. Our analysis yields an upper limit of $\alpha_{\rm stat,95\%} < 1.14$ for the combined {\it Fermi}-LAT and LST-1 dataset and $\alpha_{\rm stat,95\%} < 2.33$ for the combined {\it Fermi}-LAT and MAGIC dataset, both at the $95\%$ confidence level, without accounting for systematic uncertainties, discussed in Appendix \ref{sec:systematics}.

The upper limits on the EBL scaling factor $\alpha$, derived through the likelihood ratio test technique, can be finally turned into constraints on the EBL density, as shown in Fig.~\ref{ebl_res}. This figure represents the upper limits obtained for both the \textit{Fermi}-LAT+MAGIC and \textit{Fermi}-LAT+LST-1 spectra, {using the EBL intensity predicted} by \citet{saldana-lopez2021} as a template. The constraints on the EBL wavelength range were determined by decomposing the spectrum of the same EBL model template into discrete wavelength bins.
Each bin was treated as an independent component, and its normalization was varied systematically to assess its contribution to the total likelihood. We employed two complementary methods, following the approach introduced by \citet{pks1441_magic}: first, the normalization of a single bin was varied to induce a change of \(\Delta \chi^2 = 1\) in the likelihood ratio; second, the global \(\chi^2\) was evaluated by independently fixing the normalization of each bin to the global upper limit. As a result of this analysis, we determined that our dataset provides the most stringent constraints on the EBL within the wavelength range of approximately \(0.325 \, \mu\mathrm{m}\) to \(1.069 \, \mu\mathrm{m}\), a region that lies squarely within what is commonly known as the Cosmic Optical Background (COB). At a wavelength of \(0.6 \, \mu\mathrm{m}\), we derived upper limits (statistical, $95\,\%$ confidence level) of \(< 6.72 \) nW m\(^{-2}\) sr\(^{-1}\) with LST-1 and \(< 13.7 \) nW m\(^{-2}\) sr\(^{-1}\) with MAGIC. Adopting a conservative systematic error of up to $\pm 15\%$ in energy scale of the corresponding ground detector, the limits become \(< 8.74 \) nW m\(^{-2}\) sr\(^{-1}\) for LST-1 and \(< 14.7 \) nW m\(^{-2}\) sr\(^{-1}\) for MAGIC (see Appendices \ref{sec:systematics} and \ref{sec:ebl_profiles} for further details).

\begin{table*}[htp]
\caption{Results of the joint LAT plus LST-1 and LAT plus MAGIC spectral fits for the December 2023 flare.}
\label{tab:sedfit}
\centering
\begin{tabular}{c c c c c}
\hline\hline 
\multirow{2}{*}{Name}   &$\phi_{\rm 0}(E_{\rm 0})$  & \multirow{2}{*}{$\alpha$} & \multirow{2}{*}{$\beta$} & $E_{\rm peak}$  \\
 &   [$10^{-9}\,{\rm cm^{-2}}\,{\rm s^{-1}}\,{\rm TeV^{-1}}$]    &   && [GeV]  \\\hline
\hline
LST-1 December &  $11.0 \pm 1.4 $&$2.47\pm0.11 $&$ 0.10  \pm 0.02  $&  $8.90 $\\
    MAGIC December &  $ 4.11\pm 0.56 $&$2.89\pm0.11 $&$ 0.14 \pm 0.02   $&  $ 4.53$\\\hline 
\end{tabular}
\tablefoot{Results are based on the datasets of Table \ref{table:Fermi} and adding the corresponding VHE data from Table \ref{table:VHE_analysis}. The reference energy is set in this case to 100 GeV.}
\end{table*}

\subsection{Multi-wavelength variability}\label{sec4.2}

Figure \ref{fig:MWL_LCs} shows all the datasets compiled from the different instruments, from VHE $\gamma$-rays with LST-1 and MAGIC to radio wavelengths. The VHE $\gamma$-ray emission observed by LST-1 during December 2023 was estimated at an average level of 0.3~C.U above 100 GeV ($\sim 3.7 \times 10^{-12}$~erg~cm$^{-2}$~s$^{-1}$) during the observing campaign, with however no significant night-wise variability. In January 2024, during the non-detected low emission state, we computed 95\% confidence level upper limits to the VHE emission of 0.02 C.U., and an energy flux density of $\sim 2.5 \times 10^{-12}$~erg~cm$^{-2}$~s$^{-1}$ as shown in the light curve. 
We note that for some of the nights with MAGIC and LST-1 data, we observe a mismatch between both instruments, especially for the last two nights of the observations from December 2023.
This discrepancy was carefully evaluated, being attributed to systematic effects and different performance of the instruments. Their effect is particularly important at low energies, precisely where the bulk of the emission is detected, due to the very soft spectrum of OP 313. A careful and detailed evaluation of the systematic effects for each telescope is detailed in Appendix~\ref{sec:systematics}.

The $\gamma$-ray emission as measured by \textit{Fermi}-LAT shows a flux increase of a factor $\approx50$ with respect to the average emission reported in the 4FGL catalog \citep{4fgl}. Moreover, from the representation of the spectral index, we also observe a hint of hardening of the $\gamma$-ray spectrum, with spectral indices of $<$2 during the highest activity stages (an index of $2.23 \pm 0.01$ is reported in the 4FGL-DR4 catalog).

At X-ray wavelengths, we observe the energy flux changing from $\sim$$5.9 \times 10^{-12}$~erg~cm$^{-2}$~s$^{-1}$ and $\sim$$3.9 \times 10^{-12}$~erg~cm$^{-2}$~s$^{-1}$ at the moment of maximum emission for the soft and hard X-ray bands, respectively, to $\sim$$1.3 \times 10^{-12}$~erg~cm$^{-2}$~s$^{-1}$ and $\sim$$7.6 \times 10^{-13}$~erg~cm$^{-2}$~s$^{-1}$ when the source is faintest. As shown in Fig.~\ref{fig:MWL_LCs}, the amplitude of the variations is larger in the soft X-ray band (0.3-2~keV), whereas the peak is less pronounced in the hard X-ray band (2-10~keV). The energy flux in the different optical-UV filters varies from $\gtrsim 0.4 - 1.0 \times 10^{-11}$~erg~cm$^{-2}$~s$^{-1}$ at the peak of the flare, depending on the filter, to roughly a factor 10 lower during January 2024. As a comparison, the long-term study performed in \cite{magic2024} on several FSRQs, including OP 313, shows an average $R$-band emission of this source at a level of $\sim$0.2~mJy, equivalent to an energy flux of $\sim 1.0 \times 10^{-12}$~erg~cm$^{-2}$~s$^{-1}$, with a maximum of $\sim$2.4~mJy ($\sim 1.0 \times 10^{-11}$~erg~cm$^{-2}$~s$^{-1}$) during a flare occurring in 2019, when the source was not detected by the MAGIC telescopes. Therefore, this flare shows a maximum emission $\sim$10 times higher than the average observed in the historical data presented in \cite{magic2024}, and roughly at the level of the maximum observed in 2019.
Finally, the radio emission is approximately stable at flux density levels of $\sim$ 2 to 3 Jy depending on the band. This can be observed {in} Fig.~\ref{fig:MWL_LCs}, and was also confirmed by a continuation of the radio observations after January 2024. The data immediately after January 2024, that will be included in a future dedicated publication as they lie beyond the time period considered here, were found to be at the same flux density level as the data presented in this work (private communication).

Visual inspection of the light curves at different wavelengths shows that the emission of the different bands presents a high degree of correlation. Beyond the radio frequencies, the flaring state from December 2023 is present simultaneously in all bands, followed by the same low and relatively constant emission state in January 2024. The radio fluxes display a rather constant behaviour over time, although they have the poorest coverage among all datasets. This correlated behaviour was already observed for OP~313 by \cite{pandey2024} between its optical and $\gamma$-ray emission for another major flare occurring in 2022.

A strong indication of correlated variability is notably observed between the UV and X-rays. Between those bands, the Pearson's R coefficient is 0.93 with a significance of $5.4\sigma$ \citep[computed using the prescription of][]{press2007numerical}. {Evidence} of correlation is also found between the 0.3-2\,keV band and the \textit{Fermi}-LAT fluxes, yielding a Pearson's R coefficient of 0.85 and a significance of $4.6\sigma$ \citep[also derived following][]{press2007numerical}. The corresponding flux-flux plots are shown in Sect. \ref{sec:breakdown_model_mwl_correlations} and Fig. \ref{fig:uv_xray_corr}. We searched for time lags between the UV, X-ray and \textit{Fermi}-LAT light curves making use of the discrete correlation coefficient \citep[DCF;][]{Edelson1988}, but the results are consistent with a correlation occuring with no time lag. 

The development of the flare is accompanied by a change of the EVPA, measured thanks to the polarimetric data taken during this period. The polarization angle was measured to be $\sim -30^{\circ}$ roughly at the moment of maximum emission, swinging clock-wise approximately 90$^{\circ}$ during the decay of the flare, up to an orientation of $\sim$60$^{\circ}$. 

\subsection{Broadband spectral energy distribution}\label{sec4.3}
With the multi-wavelength data compiled for the flaring and low emissions states of December 2023 and January 2024, respectively, we characterise the broadband emission and SED of OP~313 in each state within common blazar radiative scenarios, with the aim of understanding the physical processes involved and responsible for the observed flare. The models are shown in Fig.~\ref{fig:MWL_SED} and the parameters describing each of them are included in Tables~\ref{tab:parameters_model_mireia} to \ref{tab:parameters_model_axel}. For the SED model of the flaring state, and with the aim of avoid biasing the model due to the large variability shown by the source at X-ray, UV and optical wavelengths during the complete LST-1 and MAGIC observing windows, we have selected a subperiod with a relatively stable emission, ranging from December 11\textsuperscript{th} 12:00 UTC (MJD 60289.5) to 17\textsuperscript{th} 12:00 UTC (MJD 60295.5). For the characterisation of the low state, owing to the relatively constant emission during the whole period, we consider all data between January 8\textsuperscript{th} 12:00 UTC (MJD 60317.5) and 22\textsuperscript{nd} 12:00 UTC (MJD 60331.5), covering all multi-wavelength data taken between the first and last LST-1 and MAGIC observations in this period. We note that during the time interval considered for the low state, there is no simultaneous radio observation, the closest being performed on January 2\textsuperscript{nd} and 24\textsuperscript{th}. However, as observed in Fig.~\ref{fig:MWL_LCs} and discussed in Sect.~\ref{sec4.2}, the radio emission is constant within errors. Therefore, we include the observation from January 2\textsuperscript{nd} within the data used for the SED modelling, as it lies within the time period  studied here. The stability of the radio emission is also illustrated in Fig.~\ref{fig:MWL_SED}, where the radio data used here for December 2023 and January 2024 are at the same energy flux level, and also at the average emission state when compared with the historical data of this source.

Here we have used version 0.4.0 of the \textsc{python} package \texttt{agnpy} developed for modelling the radiative processes of jetted AGNs \citep{nigro2022,nigro2023}, describing the broadband SEDs within the leptonic framework based on two emitting regions located at different distances from the central engine. It is well established that single-zone SSC models usually fail to reproduce the entire broadband SED of blazars, in particular in the radio band where several regions in the jet can contribute significantly to the observed flux \citep[see e.g.][]{pushkarev_2010}. A single-zone leptonic model was also found to be unsuccessful when trying to reproduce the broadband emission of OP~313, and at least two emitting regions are needed. Therefore, we consider a two-zone leptonic scenario where the radiation produced by each emitting region is characterized by a set of physical parameters including its radius $R$, magnetic field $B$, bulk Lorentz factor $\Gamma$ and Doppler factor $\delta$. The population of relativistic electrons responsible for the emission in each region is described by a broken power-law function with the Lorentz factor of the particles $\gamma$ as 

\begin{equation}
N(\gamma) = \left\{ \begin{array}{lc} K(\gamma/\gamma_{b})^{-n_{1}}: & \gamma_{min}<\gamma<\gamma_{b} \\ \\  K(\gamma/\gamma_{b})^{-n_{2}}: & \gamma_{b}>\gamma>\gamma_{max}  \end{array} \right.
\label{eq:broken_PL_electrons}
\end{equation}
where $K$ is the normalization of the distribution between the minimum and maximum values of the Lorentz factor, $\gamma_{min}$ and $\gamma_{max}$, and $n_{1}$ and $n_{2}$ are the spectral indices of the distribution below and above the break $\gamma_{b}$, respectively.

We have also included in the model the contributions from the BLR, DT and  AD to the infrared-to-UV emission, as well as the $\gamma$-ray emission via EC from low-energy photons provided to the jet by these components. First, we have used the optical spectrum from the Sloan Digital Sky Survey (SDSS)\footnote{\url{https://www.sdss.org}}, were a prominent broad MgII emission line is visible at 5590~\AA~\citep[see][]{schneider2010}, to estimate the luminosity of the BLR. The amplitude of this emission line was estimated to be $1.33\times 10^{-16}$~erg~cm$^{-2}$~s$^{-1}$~\AA$^{-1}$, equivalent to a flux of $1.02 \times 10^{-14}$~erg~cm$^{-2}$~s$^{-1}$. Adopting the cosmological parameters $H_{0}=69.6$~km~s$^{-1}$~Mpc$^{-1}$, $\Omega_{m}=0.286$ and $\Omega_{\Lambda}=0.714$ reported by \citet{Bennett_2014}, we obtain a luminosity distance $d_{L}=6680$~Mpc that allows us to calculate the luminosity of the MgII line, resulting in a value of $L_{\rm MgII}=5.44 \times 10^{43}$~erg~s$^{-1}$. We can then calculate the total luminosity of the BLR with using the relations between the different contributions of the emission lines to the total luminosity of the BLR reported by \citet{ghisellini2015} and \citet{finke2016}. According to these authors, the MgII emission corresponds to 34/556 of the total BLR luminosity, which results in $L_{\rm BLR}\sim9.0\times 10^{44}$~erg~s$^{-1}$. Then, using the approximation between the AD and BLR luminosities as $L_{\rm  AD}\simeq10\times L_{\rm BLR}$, we obtain a disk luminosity of $L_{AD} \sim 9.0 \times 10^{45}$~erg~s$^{-1}$.
Finally, with the relation reported by \citet{vestergaard2009} between $L_{\rm MgII}$ and the black hole mass $M_{\rm BH}$, we constrain the latter to a value of $M_{\rm BH} \sim 4 \times 10^{8}M_{\odot}$. The models presented here were fitted {in an empirical manner}, with no minimization algorithm used.

\subsubsection{Spectral-focused model}\label{sec:model_balanced}

\begin{figure}[ht!]
\centering
\includegraphics[width=\columnwidth]{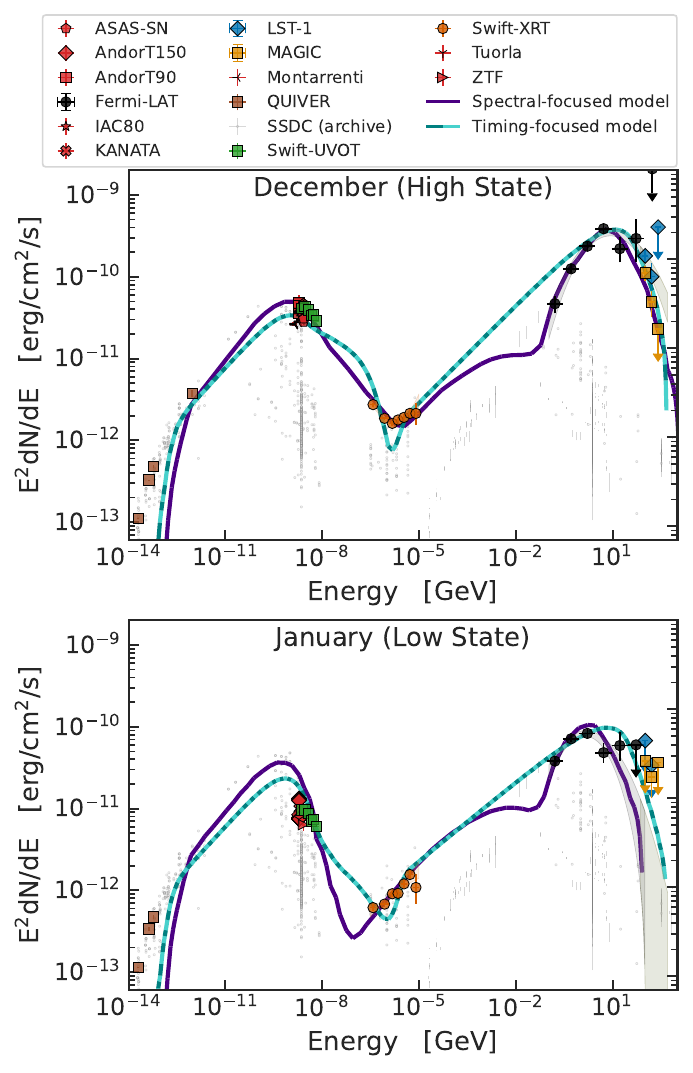}
\caption{\label{fig:MWL_SED}Broadband SED models of OP 313. \textit{Top:} Broadband SED during the December 2023 flare. \textit{Bottom:} Broadband SED during the January 2024 low state. Different markers represent data at different wavelengths as indicated in the legend. Purple solid line and light blue dashed lines represent the spectral-focused model of section \ref{sec:model_balanced} and the timing-correlation model from section \ref{sec:model_mwl_correlations} respectively. Grey markers correspond to archival data extracted from the SSDC database (\url{https://tools.ssdc.asi.it/SED/}). All flux points have been corrected for extinction, hydrogen absorption and EBL attenuation. }
\end{figure}

The first model, represented in the top panel of Fig.~\ref{fig:MWL_SED}, stems from the {balanced} contribution of two external photon fields to the $\gamma$-ray emission, this is, BLR and DT, and aims to carefully reproduce the observed SED in the optical-UV, X-ray and $\gamma$-ray bands, including the observed spectral curvature in each band and its evolution. 
The jet consists {of} two non-interacting populations of relativistic electrons: a colder, stable, and more extended ``far zone'' region (radius $R'_{\rm far}=1.5\times10^{17}\,$cm and {placed at a distance from the black hole} $R_{H,\rm far}=5\times 10^{19}\,$cm)  responsible for the low energy component (IR, optical) and dominates in the X-rays above $\sim1$\,keV; and a much more energetic ``near zone'' (radius $R'_{\rm near}=1\times10^{16}\,$cm and located at a distance from the black hole $R_{H,\rm near}=2.9\times 10^{17}\,$cm), which is beyond but not far from the BLR, and thus collects a significant external Compton contribution on the BLR field and the DT. The absorption effects due to the external photon fields are included in the model. In this scenario, absorption effects are dominated by the DT field and occur above a few hundreds of GeV.
This ``near zone'' is dynamic in particle density and energy, and fully capable of enhancing the low energy X-ray emission during the high states (see Figure \ref{fig:uv_xray_corr}). It also dominates completely through external Compton the $\gamma$-ray band, both during low and high states. For the external photon fields, we assume {an accretion efficiency $\eta = 1/12$ \citep[consistent with a Shakura \& Sunyaev disk model, see][]{Shakura_Sunyaev_1973}}, a covering factor of the DT $\Xi_{\rm DT}=0.2$, corresponding to a DT luminosity of $L_{\rm DT} \sim \Xi_{\rm DT} \times L_{\rm AD} = 1.8 \times 10^{45}$~erg~s$^{-1}$, and a reprocessing factor of the BLR (based on the ${\rm Ly \alpha}$ line) $\xi_{\rm Ly \alpha}=0.023$. {The latter values are similar to the estimates reported in \citet{finke2016}}. All the parameters describing this model are summarized in Table~\ref{tab:parameters_model_mireia} and \ref{tab:thermal_components}. The model assumes an exceptionally high Doppler factor of $\delta = 99$ for the ``near zone'' to simultaneously {capture the SED from the X-ray to the} $\gamma$-ray bands, as well as the V-shaped X-ray spectrum observed during the high state. {This should be regarded as one possible solution, guided by the observed spectral shape, and not necessarily a unique one given the known degeneracies of leptonic modeling. While such extreme values are not directly supported by observational constraints from Very Long Baseline Interferometry (VLBI) for this source, they cannot be ruled out. Similarly high Doppler factors have been invoked to explain fast VHE flares in other FSRQs, such as PKS~1510-089 or 3C~279 \citep{2016MNRAS.463L..26B, 2020NatCo..11.4176S}}. The model excels at explaining the evolution of the SED from $\sim 10^{11}\,$Hz to the VHE band, requiring only changes in the population of electrons for the ``near zone'', and overshooting the recorded optical/UV emission during the low state. The resulting models for December and January are shown in Fig.~\ref{fig:MWL_SED}, and in more detail in Fig.~\ref{fig:MWL_SED_model_balanced}.\par 
{The stable physical properties of the emitting region (for instance the magnetic field $B'$ or the radius $R'$) that we obtain to describe the emission over monthly timescale points towards an emitting region located behind a stationary shock within} the jet. Indeed, if the emission were to originate from a blob travelling downstream the jet (as in the blob-in-jet model), the large $\delta$ would imply a distance travelled by the blob of $\approx 1$\,kpc during $\approx1$\,month (assuming a bulk Lorentz factor of 50). Over such a large distance, the physical properties of the emitting region are expected to vary significantly, and the target photon field density for the external Compton scattering would drastically decrease. The flare may therefore originate from stationary jet features, such as recollimation shocks \citep[see e.g.][]{hervet2019}. This argument also motivated our choice to fix $R_{H}$ between the epochs. We stress that even if a stationary shock is considered, the luminosity in the observer's frame is boosted with the usual $\delta^4$ scaling \citep{1997ApJ...484..108S}.

\begin{table}
\caption{\label{tab:parameters_model_mireia} Parameters of the spectral-focused emission model of section \ref{sec:model_balanced}.}
\centering
\resizebox{\columnwidth}{!}{%
\begin{tabular}{l c c c @{\hskip 0.3in} c}   
\hline\hline
\multirow{2}{*}{Parameters} & \multicolumn{2}{c}{``near zone''} &  & \multirow{2}{*}{``far zone''} \\\cline{2-3}
  & December 2023 & January 2024 &   &  \\
\hline
\hline
$B'$ [$10^{-2}$ G]           & 45     & 45    & & 30 \\
$R'$ [$10^{16}$ cm]          & 1      & 1     & & 15 \\
$\delta$                    & 99     & 99    & & 19 \\
$\Gamma$                    & 50     & 50    & & 10 \\
$N_e$ [$10^{-5}$ cm$^{-3}$] & $29$   & $100$   & & $20$ \\
$n_1$                       & $1.8$  & $1.8$ & & $2.1$ \\
$n_2$                       & $4.0$  & $4.0$ & & $--$ \\
$\gamma'_{min}$ & $3\times10^{2}$ & $3\times10^{2}$ & & $15$ \\
$\gamma'_{br}$  & $2.5\times10^{3}$ & $1\times10^{3}$ & & $--$ \\
$\gamma'_{max}$ & $5\times10^{4}$ & $7\times10^{3}$ & & $ 6\times10^{3}$ \\
$R_H$ [$10^{17}$ cm] & 2.9 & 2.9 & & 500 \\
\hline
\hline
\end{tabular}
}
\tablefoot{ Parameters are obtained with comparable contributions of the DT and the BLR to the total external Compton emission.
\tablefoottext{a}{We refer the reader to Sect.~\ref{sec4.3} for a detailed description of each parameter. 
}
}
\end{table}

\begin{table}
\caption{\label{tab:thermal_components} Parameters for the thermal components of OP~313 (AD, DT, BLR).} 
\centering
\begin{tabular}{l c}    
\hline\hline
Parameters & Values \\
\hline
\hline
$M_{\rm BH}$ [$M_\odot$] & $10^{8.57}$ \\
$L_{\rm AD}$ [$10^{45}$ erg s$^{-1}$] & 9 \\
$\eta$ & 1/12 \\
$R_{in}$ [$R_g$] & 5 \\
$R_{out}$ [$R_g$] & 50 \\
$L_{\rm BLR}$ [$10^{44}$ erg s$^{-1}$] & 9 \\
$\xi_{Ly \alpha}$ & 0.023 \\
$R_{Ly \alpha}$ [$10^{16}$cm] & 9.5 \\
$L_{\rm DT}$ [$10^{45}$ erg s$^{-1}$] & 1.8 \\
$T_{\rm DT}$ [K] & 800\\
$\Xi_{\rm DT}$ & 0.2 \\
$R_{\rm DT}$\tablefootmark{a} [$10^{16}$cm] & 1875 \\
\hline
\hline
\end{tabular}
\tablefoot{Parameters are shared among both the spectral-focused emission model and the timing-focused emission model and constant over time.
\tablefoottext{a}{$R_{DT}$ is not a parameter of the model, but is calculated as the sublimation radius following Eq.\,(96) of \cite{finke2016}.}
}
\end{table}

\subsubsection{Timing-correlation model}\label{sec:model_mwl_correlations}
We investigated a second scenario that aims to qualitatively reproduce the observed multi-wavelength correlation patterns (see Sect.~\ref{sec4.2}) while adopting a less extreme Doppler factor with respect to the spectral-focused model. The correlation between UV, X-ray and $\gamma$-ray fluxes suggests a common underlying particle population responsible for the optical-to-VHE flaring activity. We therefore propose a modelling scenario for the December 2023 and January 2024 epochs where a single emitting zone (that we also dubbed as the ``near zone'') dominates {entirely} the emission beyond the optical/UV band. At lower frequencies, in the radio regime, the emission originates from a separate, broader region called the ``far zone''. In the ``near zone'', the electron distribution is modelled with a broken power-law as defined in Eq.~(\ref{eq:broken_PL_electrons}). For the ``far zone'', instead, a simple power-law function is used, $N(\gamma)=K\gamma^{-n_1}$. This reduces the number of free parameters, while preserving a satisfactory description of the data. Furthermore, the optical-to-radio data do not allow to strongly constrain the additional parameters introduced by the broken power-law model.\par

We use the same parameters as in Model 1 to describe the BLR, AD and DT. To avoid significant absorption of the $\gamma$-ray photons by the BLR field and to reproduce the spectrum up to VHE, the ``near zone'' is located at a distance of $1.9\times10^{17}$\,cm away from the central engine, slightly beyond the BLR that has a radius of $9.5\times10^{16}$\,cm. In such configuration, the dominant target photon field for inverse-Compton scattering is still the one from the BLR, and it dominates over the DT radiation. The ``far zone'' is placed even further downstream, at $2\times10^{19}$\,cm, such that the resulting inverse-Compton emission on the DT field remains sub-dominant in the entire $\gamma$-ray band with respect to the ``near zone''. Finally, we kept the model parameters in the ``far zone'' constant between the December 2023 and January 2024 epochs. This choice is motivated by the low variability in the radio bands. Regarding the ``near zone'', we only evolve the parameters related to the electron distribution. The resulting models for both epochs are shown in Fig.~\ref{fig:MWL_SED}, and in more detail in Fig.~\ref{fig:MWL_SED_model_mwlcorr}. The contribution from the ``near zone'' and ``far zone'' are plotted with thick and thin solid lines respectively. We list in Tables~\ref{tab:thermal_components}~and~\ref{tab:parameters_model_axel} the corresponding parameters.\par

\begin{table}
\caption{\label{tab:parameters_model_axel}Parameters of the timing-correlation model that focuses on reproducing the multi-wavelength correlation pattern of Sect.~\ref{sec:model_mwl_correlations}}.
\centering
\resizebox{\columnwidth}{!}{
\begin{tabular}{l c c c @{\hskip 0.3in} c}    
\hline\hline
\multirow{2}{*}{Parameters} & \multicolumn{2}{c}{``near zone''} &  &  \multirow{2}{*}{``far zone''} \\\cline{2-3} 
  & December 2023 & January 2024 & \\
\hline
\hline
$B'$ [$10^{-2}$ G]  & 45 & 45 & & 45 \\
$R'$ [$10^{17}$ cm] & 1 & 1 & & 5 \\
$\delta$ & 50 & 50 & & 20\\
$\Gamma$ & 25.3 & 25.3 & & 10\\
$N_e$ [cm$^{-3}$] & $5.4$ & $3.8$  &  & 11.1\\
$n_1$ & 2.2 & 2.4 & & 2.1 \\
$n_2$ & 3.4 & 3.7 & & -- \\
$\gamma'_{min}$ & 1 & 1 & & 1\\
$\gamma'_{br}$ & $5\times10^{3}$ & $4\times10^{3}$ & & -- \\
$\gamma'_{max}$ & $4\times10^{4}$ & $5\times10^{4}$ & & $5\times10^{3}$\\
$R_H$ [$10^{17}$ cm] & 1.9 & 1.9 & & 200 \\
\hline
\hline
\end{tabular}
}
\tablefoot{ We refer the reader to Sect.~\ref{sec4.3} for a detailed description of each parameter. }
\end{table}

\section{Discussion}\label{sec5}

\subsection{Multi-wavelength evolution}

The flare of OP\,313 in December 2023 shows increased flux levels in almost all multi-wavelength bands (see Fig.~\ref{fig:MWL_LCs}). This is also evidenced by the high correlations reported in Sect.~\ref{sec4.2} between UV and X-rays (5.4$\sigma$) and between the X-rays and high-energy $\gamma$-rays ($4.6\sigma$). For the VHE regime, the poorer time coverage does not allow to identify correlations with other energy bands. Nonetheless, this indicates that the December 2023 flare is originated in a single emission zone that dominates the emission from optical-UV to $\gamma$-rays.

In addition, as reported by \cite{otero2023}, during a first flare happening in November 2023, prior to the detection by LST-1 and MAGIC, and during which the IACTs could not {observe} due to the full moon period, a remarkably high optical polarization degree up to $\sim$22\% was also observed. Owing to the scarce coverage during December, we cannot claim the same behaviour for this flare where the polarization measurements during the flare show polarization degrees around and below 10\%.
However, the optical polarization angle depicts a swing of $\sim $90$^\circ$ in December 2023 during the flare. Such angle rotations have been observed for other FSRQs before and are often accompanied by multi-wavelength flaring activity \citep[see e.g.][]{2011A&A...530A...4A,kiehlmann2016,2017A&A...603A..29A,2018A&A...619A..45M}. Nevertheless, we note that in this case, the poor sampling prevents us from deriving stronger conclusions about the nature of this event. The radio polarization angles increases slightly towards the end of the flare, but does not show strong variability. 
Variability in the polarization angle has been observed before for OP\,313 \citep{2017A&A...602A..29B} and were considered as natural consequences of helical magnetic fields inside the blazar's jet.

\subsection{$\gamma$-ray SED and EBL constraints}\label{sec:sed_ebl_discussion}

In the \textit{Fermi} catalog, the mean power-law spectral index of the FSRQ population is $2.44\pm0.20$ \citep{2020ApJ...892..105A}. Knowing that the majority of FSRQs spectra {are} well {fitted} with a power-law (446 over 694), this means that  the bulk of FSRQs detected in HE have a peak position below the \fla\ energy range. In the 4FGL catalogue \citep{4fgl}, OP 313 spectra is well fitted by a log-parabola and exhibit a peak position $\sim 0.5$\,GeV.

We can evaluate the peak position during the December flare using the joint LAT-VHE fit to be around a few GeV (see Table~\ref{tab:sedfit}). This fit shows that the peak position has shifted towards a higher energy during this event. Other FSRQs detected during outburst states \citep{2011ApJ...730L...8A,pks1441_veritas,2021A&A...647A.163M} also present a high peak position, PKS~1510-089 being one of the most striking cases with a peak position between 10 and 60~GeV during a flare in 2016 \citep{2021A&A...648A..23H}.

\begin{figure}[ht!]
    \centering
        \centering
        \includegraphics[width=\columnwidth]{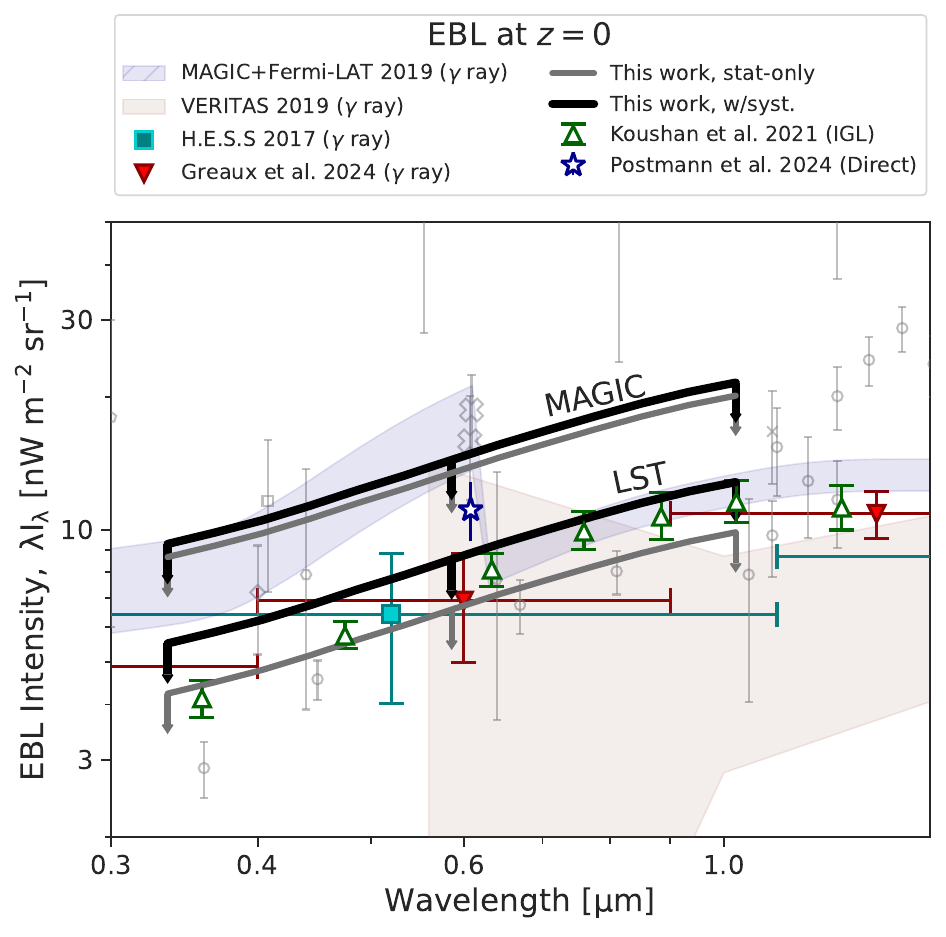}
    \caption{EBL intensity as a function of wavelength at \(z = 0\). Indirect constraints derived from $\gamma$-ray observations are shown as follows: MAGIC+\textit{Fermi}-LAT results by \citet{2019MNRAS.486.4233A} (light blue hatched region), VERITAS results by \citet{Abeysekara_2019} (light brown region), H.E.S.S. results by \citet{2017A&A...606A..59H} (turquoise square points), {and archival MAGIC, H.E.S.S. and VERITAS TeV spectra by \citet{Greaux_2024} (red triangles)}. Direct measurements from \citet{Postman_2024} correspond to the dark blue star markers and IGL estimates from \citet{Koushan_2021} are represented with the green triangles. 
    }
    \label{ebl_res}
\end{figure}

The EBL constraints obtained in Sect. \ref{sec:sed_ebl} are well in agreement with previous indirect measurements by MAGIC \citep{2019MNRAS.486.4233A}, VERITAS \citep{Abeysekara_2019} and H.E.S.S. \citep{2017A&A...606A..59H}. They also provide an independent consistency check to the latest integrated galaxy light (IGL) estimates. \citet{Koushan_2021} introduced major improvements over \citet{Driver_2016}, refining their analysis to reduce fragmentation and false detections while improving error treatment, which reduced optical/NIR IGL uncertainties from 20\% to below 10\%. Their updated IGL estimates also show a modest increase of about 5–15\% in all wavebands, {bringing} them slightly closer to our upper limits. As shown in Fig.~\ref{ebl_res}, our MAGIC and LST-1 results remain in good agreement with these latest constraints, indicating that {unknown astrophysical source classes} are unlikely to account for a significant fraction of the EBL. Nevertheless, while our most stringent limits for LST-1 are statistically consistent with the latest direct measurements from New Horizons LORRI \citep{Postman_2024} within uncertainties, this agreement remains inconclusive, as also noted by \citet{Koushan_2021} and \citet{Greaux_2024}. Consequently, the possibility of additional components contributing to the COB, whether from {new} astrophysical sources or new physics, cannot yet be definitively ruled out.

The LST-1 limits are more constraining that the ones obtained with the MAGIC data. We investigated if the longer LST-1 exposure with respect to MAGIC could explain this difference by restricting the analysis over strictly simultaneous nights. Still, no significant change was observe and the LST-1 limits remained more constraining. The difference is most probably due to a mix of higher statistics, lower energy threshold and harder spectral shape in the LST-1 data with respect to MAGIC.

\subsection{Broadband SED modelling}
We have modelled the broadband SED of OP~313 employing a leptonic scenario, accounting not only for the synchrotron and SSC emission intrinsic to the jet but also for the high-energy emission via EC scattering of the photon field from the AD, BLR and DT. OP~313 exhibits a large Compton dominance (ratio between the high- and low-energy peak of the SED; see Fig.~\ref{fig:MWL_SED}) of $\sim8$. This spectral feature is commonly observed in other FSRQs and the presence of low-energy photon fields is necessary to produce via EC scattering a $\gamma$-ray flux level compatible with the observations.

We consider a two-zone model, as single-zone scenarios were found to be insufficient to describe the broadband emission. We consider two different approaches. 
In the first model, the emission zone is located close to the BLR leading to a relatively balanced combination of EC emission due to BLR and DT photons, while the contribution from the AD is negligible. This model is able to accurately
describe the broadband emission of OP~313 except in the optical band, which would require a coherent variability of the two zones, difficult to explain since the two zones are considered to be spatially separated, therefore, strong correlations are not necessarily expected. For both zones, the size of the emitting region is in agreement with the constraints set by causality arguments based on the variability observed \citep[see e.g.][]{dondi1995}, on the order of $\sim$1-day timescales, with the more compact ``near zone'', being $\sim$12 times smaller than the maximum allowed size by the variability.
In addition, the model predicts a very large Doppler factor for the near zone. This, however, is not {very} surprising given the powerful high-energy flare that dominates the emission, where a strong boosting effect is present in a relatively compact region. The spectral indices of the electrons, 1.8 and 4.0 before after the break respectively, are in line with the expectations from shock acceleration ($2-2.5$) and magnetic reconnection (where even harder indices are possible).
Nevertheless, this scenario manages to capture the change in emission behavior for OP\,313 by invoking decreases in \(\gamma'_{\rm break}\), \(\gamma'_{\rm max}\), and the particle density \(N\) within the compact region, without modifying the larger zone.
Within this model, the underlying cause for the spectral changes is mainly rooted in the injection and cooling of the particle populations.

The second scenario is motivated by the strong multi-wavelength correlations and also aims to explore the possibility of a lower Doppler factor with respect to the spectral-focused model. It considers a single component (the ``near zone'') that explains the flaring state observed simultaneously from optical-UV to VHE $\gamma$ rays. Again, the size of this compact region is in agreement with the constraint set by the variability. A ``far zone'' only contributes significantly at radio frequencies, where no significant outburst was detected. In addition to the correlation trends, such a two-component morphology is justified by the difference in the polarization degree noticed between the radio and R-band. In the R-band, the polarization degree is at the level of $\approx10$\% throughout the campaign, although the radio polarization is only of a few~\%, implying a spatial separation of the two emitting regions (at least partial). 
Compared to the first model, the emission region is placed slightly closer to the central engine and the external Compton radiation is dominated by emission due to photons from the BLR.
This model reaches a fairly good reproduction of the broadband SED, however poorer than the previous scenario during the high-state and in particular in the optical-to-X-rays frequency range. During the flare, the VHE spectral hardness requires an electron distribution with an index of $n_2 \approx 3.4$ beyond the break, leading to a synchrotron emission in the optical-to-X-ray  band that is somewhat harder than suggested by the data. The difference between the two epochs (high and low state) is mostly explained by a significant softening of the electron distribution indices $n_1$ and $n_2$ over time. $n_1$ evolves from 2.2 to 2.4 between the December 2023 and January 2024 epochs, while $n_2$ changes from 3.4 to 3.7. Simultaneously, the $\gamma'_{\rm  max}$ drops from $5\times 10^4$ to $4\times 10^4$. 

In the latter scenario the bulk Lorentz factor adopted in the ``far zone'' ($\Gamma = 10$) is consistent with recent kinematic studies of OP~313 on parsec scale with VLBI and \citet{weaver2022} found $\Gamma_{\rm VLBI} \approx 15$. A significantly higher values is required for the ``near zone'', $\Gamma = 25$, but it remains within the typical range found in FSRQs \citep{weaver2022}. The ``near zone'' could be associated to the ejection of a fast, new plasma blob close to the base of the jet.

To evaluate the energetic requirements, we estimate the kinetic jet power as:
\begin{equation}
    L_{jet, kin} \simeq \pi \; R'^2 \; \Gamma^2 \; \beta c \; (u_{p} + u_{e} + u_{B})  \ ,
\end{equation}
where $u_e$ is the electron energy density, $u_B=B'^2/8\pi$ the magnetic energy density and $u_p \approx N_p \,m_p c^2$ is the 
(cold) proton energy density. We assume an equal number density of protons and electrons, such that $N_p = N_e$. For both epochs, the total power (obtained by summing the ``near'' and ``far'' zones) is at the level $L_{jet, kin} \simeq 9 \times 10^{46}$\,erg\,s$^{-1}$ and $L_{jet, kin} \simeq 3 \times 10^{46}$\,erg\,s$^{-1}$ for the timing-focused and spectral-focused model, respectively. This implies a luminosity comparable or slightly above the Eddington luminosity $L_{Edd} \simeq 5 \times 10^{46}$\,erg\,s$^{-1}$ for a black hole mass of $M_{\rm BH} \sim 4 \times 10^{8}M_{\odot}$, as estimated in Sect.~\ref{sec4.3}.

\section{Conclusions}\label{sec6}
We detect for the first time significant VHE $\gamma$-ray emission from the FSRQ OP~313, becoming the tenth FSRQ ever characterized as a VHE $\gamma$-ray emitter, and the furthest AGN ever observed at such high energies. This is also the first scientific discovery performed with the new generation of Cherenkov telescopes thanks to the lower energy threshold and increased sensitivity at a few GeV of LST-1. The LST-1 and MAGIC datasets, with an extensive coordinated multi-wavelength campaign, allowed a detailed study of its broadband behaviour.

Making use of the LST-1 and MAGIC data, we characterize the observed VHE $\gamma$-ray emission from the flare detected in December 2023, and the posterior low emission state from January 2024 during which the source was not detected by the IACTs. The average emission at VHE of OP~313 during the flaring state was found to be 0.3~C.U., with no significant night-wise variability detected. During the low emission state we compute upper limits of $\sim$0.5\% of the Crab Nebula flux above an energy of 100 GeV. The observed spectrum displays a very soft behaviour, as expected due to the large EBL attenuation for such a distant source. After correcting {for} this absorption, we estimate the intrinsic spectrum, with a spectral index $\alpha = 1.99 \pm 0.55$. Consistent results within uncertainties --- especially considering systematic sources of uncertainty at a few GeVs --- are observed by MAGIC.

We have also extended the $\gamma$-ray spectral analysis from $\sim$100~MeV to a few hundreds of GeV through a joint LAT-LST-1 and LAT-MAGIC fit, using strictly simultaneous data. This multi-instrument analysis enabled a detailed scan of the EBL intensity, allowing to set constraints to the strength of the EBL absorption at redshift $z=0.997$ through the attenuation of the $\gamma$-ray spectrum. Upper limits of $\lambda I_{\lambda}<6.72$~nW m$^{2}$ sr$^{-1}$ with the combination of LAT and LST-1, and $\lambda I_{\lambda}<13.7$~nW m$^{2}$ sr$^{-1}$ with LAT and MAGIC were estimated for the EBL intensity {at 0.6 $\mu$m} (\(< 8.74 \) nW m\(^{-2}\) sr\(^{-1}\) and \(< 14.7 \) nW m\(^{-2}\) sr\(^{-1}\) including systematics).

Thanks to the intense multi-wavelength monitoring, we characterize the flare and low state at all frequencies. We observe the same behaviour in all bands, with {a} highly correlated flare followed by a low emission state in January 2024, with the exception of the radio wavelengths, that show on the other hand a rather stable emission along the monitored period.

We interpret the broadband emission with a two-zone leptonic model, including the EC contribution to the \(\gamma\)-ray emission of low-energy DT and BLR photons. We propose two alternatives, one based on a balanced EC emission from each component that aims at accurately describing the broadband spectral features. The combination of SSC and EC scattering (of the DT and BLR fields) provides a good description of the spectral shape from optical to VHE \(\gamma\)-rays, but requires the region to be relatively close to the outer edge of the BLR. Our second model, focused at explaining the correlation patterns, assumes a single emitting region---farther away from the compact object---which accounts for the bulk of the emission beyond optical-UV frequencies. In this scenario, the second SED component is entirely ascribed to EC scattering of the dusty torus field. However, the constraint of having a single zone dominating beyond the optical band leads to a less accurate modelling of the \(\gamma\)-ray component during the high state. Overall, the values of the model parameters are consistent with previous works on FSRQs modelling \citep{2019A&A...627A.140A, 2021A&A...647A.163M}.

\bibliographystyle{aa}
\bibliography{biblio}

\begin{appendix} 

\section{Evaluation of systematic effects on the LST-1 and MAGIC results}
\label{sec:systematics}
We have performed an evaluation of possible sources {of} systematic effects in the analysis of the LST-1 and MAGIC data, namely the effect of the background normalization, the choice of different efficiencies for the event selection cuts and the energy scale. These effects are detailed in the sections below. These tests were done on the dataset from December 2023, when OP~313 was detected.

\subsection{Systematics on the background normalization}\label{appA.1}
As detailed in Sect.~\ref{sec3.1}, the high-level analysis of the LST-1 data was performed using 3 OFF regions at 90$^{\circ}$, 180$^{\circ}$ and 270$^{\circ}$ from the position of the source. Using the 90$^{\circ}$ and 270$^{\circ}$ OFF positions, we have measured a total number of background counts of $487702 \pm 698$ and $486203 \pm 697$, respectively, which corresponds to a relative difference between both wobbles of $\sim$0.31 $\pm$ 0.14\%. Therefore, we have evaluated the effect that a systematic {error} of $\pm$0.5\% in the background normalization {would introduce} in our analysis, namely on the integral flux calculations. This estimation was performed with the standard 70\% efficiency cuts in gammaness and $\theta$.
We observe that the effect of fluctuations on the background estimation and normalization introduces a systematic uncertainty on the estimated flux above 100 GeV of 30\%.

For the MAGIC data, we compared the background counts between all three OFF regions and found a fluctuation of around 1\%, which is in line with the typical performance of the telescope \citep[see][]{2016APh....72...76A}. Taking into account this effect in the analysis, we find a systematic uncertainty on the integrated flux ($>$100\,GeV) of {20-30\%} percent. 

\subsection{Systematics on the event selection cut efficiency}\label{appA.2}
We have also tested the effect of different choices of the gammaness and $\theta$ selection cuts in the LST-1 data {analysis}. The analysis presented along the manuscript uses standard values of 70\% for both cuts. Here we have tested systematic effect of using limit values of 50\% and 90\% for the gammaness cut and 70\% and 90\% for the $\theta$ cut. A re-analysis of the LST-1 data with this extreme set of event selection cuts resulted in {an} estimation of a systematic source of uncertainty of 23\% of the flux above 100 GeV. 

{A similar evaluation of the systematic effect of cut selection was performed for the MAGIC data. We varied the efficiency between 70\% and 95\% and 50\% to 95\% for the $\theta$ and gammaness cuts, respectively. We then compared the outcome against the standard values used in the analysis detailed in Sect.~\ref{sec3.2} (i.e. 75\% and 90\% efficiency for theta and gammaness cuts)}. The estimated effect on the integrated flux above 100~GeV was calculated to be {on average around $\sim$15\%} \citep[see also][]{2016APh....72...76A}. 
{In addition, we evaluated the impact of lowering the minimum effective area threshold (relative to its maximum) from $5\%$ in the case of LST-1 to $1\%$ for MAGIC. The effect on the integrated flux above $100\,\mathrm{GeV}$ is below $1\%$, but becomes more significant for other threshold energy choices, reaching $\sim16\%$ for $E > 80\,\mathrm{GeV}$ and $\sim12\%$ for $E > 120\,\mathrm{GeV}$ for example.} 

We note that in both cases the evaluation of the systematics due to the cut efficiency was performed with OP~313 data. Therefore, they are correlated with the systematic effects caused by background normalization changes, reported in Appendix~\ref{appA.1}. This is because the strength of gammaness and $\theta$ cuts has a strong influence on the signal to background ratio measured. In addition, we highlight that the values reported in this section are rather conservative. {Part of the variations observed in the systematics tests likely originate from statistical variations} that are especially relevant for weak sources {for IACTs, such as OP~313}.

\subsection{Systematics on the energy scale}\label{appA.3}
A central source of systematic uncertainties in IACT observations is the one related to the absolute energy scale. Since the energy of $\gamma$-ray events {is} reconstructed based on MC simulations, a discrepancy between the simulations and the actual atmospheric transparency (or the light collection efficiency), inevitably leads to systematics in the estimated energy of the events, and in turn on the estimated flux.

This effect is exacerbated in the case of OP~313, since the bulk of the emission lies at energies where both LST-1 and MAGIC exhibit significant reconstructed energy bias. By selecting a sample of $\gamma$-ray-like events detected by both instruments, we find differences in reconstructed energies at the lowest energies of this specific dataset of up to $20-30\%$, 
consistent with a conservative estimate of $\sim 15\%$ uncertainty on the energy scale of each instrument \cite{2023A&A...680A..66A}. The effect of such energy scale systematics is evaluated below for LST-1 and MAGIC individually.

For the evaluation of systematic effects stemming from a possible bias on the absolute energy scale in the LST-1 data analysis, we introduce a $\pm15\%$ shift on the energy scale.
The energy dispersion matrix is modified by shifting it along the true energy axis, to compensate (approximately) for the possible reconstruction bias from MC-data mismatch. We then re-compute the energy spectrum and integrated flux above 100GeV.
 
We find that the assumed shift in the energy scale contributes a systematic uncertainty of about $+57\%,\ -47\%$ for the estimated fluxes above 100~GeV, and $-0.31, +0.44$ in the spectral index at $100\,$GeV. Similarly, the upper limit on the EBL scale parameter, $\alpha$, degrades to $<1.48$ when this systematic uncertainty is incorporated in the joint fit of {\it Fermi}-LAT and LST-1 data.

The uncertainty on the MAGIC fluxes resulting from systematics in the energy scale was evaluated in a similar fashion as in \citet{2019MNRAS.486.4233A}. We shifted the energy scale of the IRF by $\pm 15\%$, {corresponding to} the expected {worst-case} systematic uncertainty on the energy scale of the MAGIC telescopes \citep{2016APh....72...76A}. We then fitted the spectra, and integrated them above 100\,GeV to estimate the effect in the lightcurves shown throughout this work. 
This results on a systematic uncertainty of about $+74\%,\ -52\%$ for the fluxes above $100\,$GeV, and $-0.32, +0.16$ in the spectral index at $100\,$GeV.
Among the different sources of systematics investigated here, this is the dominant one for the MAGIC data. The corresponding EBL scale parameter upper limit, once an energy shift of $\pm 15\%$ is applied, becomes $<2.50$ for the combination of {\it Fermi}-LAT and MAGIC observations. 

\section{EBL likelihood profiles}\label{sec:ebl_profiles}

This appendix provides details on the determination of the upper limits on the EBL density. As discussed in in \ref{sec:sed_ebl} and \ref{sec:sed_ebl_discussion}, this estimation is performed by means of a likelihood ratio test, tracing the changes of the $-2\,$log-likelihood as a function of an assumed global optical depth scaling factor $\alpha$ that we scale our target EBL model with.

\begin{figure}[h!]
\centering 
\includegraphics[width=1\linewidth]{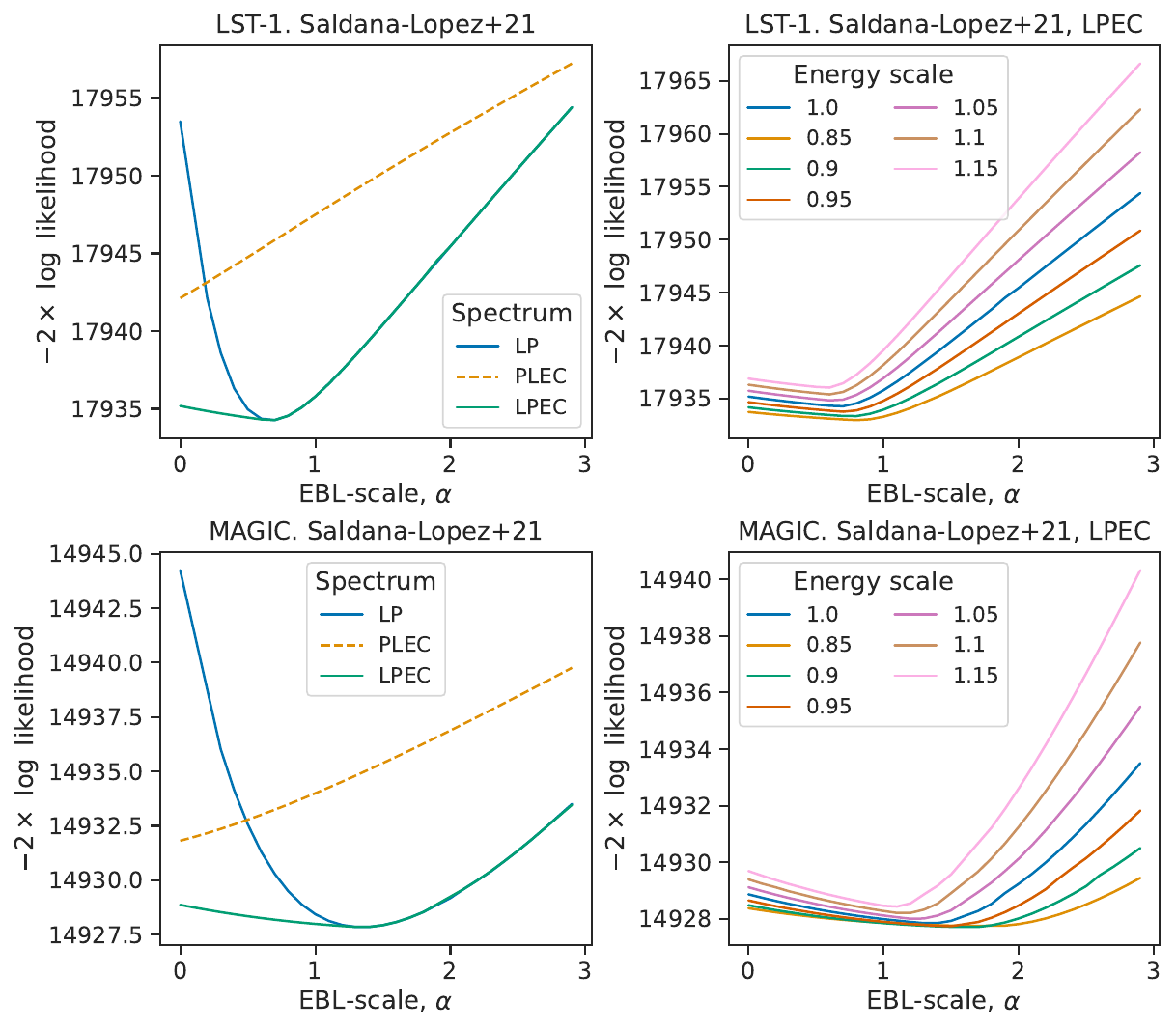}
\caption{Likelihood profiles for different assumed spectral shapes (LP, PLEC and LPEC) and for different energy scale factors (i.e. systematics) on the corresponding VHE data.}
\label{fig:EBL_Likelihood_profiles}
\end{figure}

Figure~\ref{fig:EBL_Likelihood_profiles} shows the profile likelihoods, 
\(\Delta(-2\ln \mathcal{L})\), as a function of the EBL–scale factor \(\alpha\). From top left to bottom right, panels (a)–(d) illustrate two classes of systematic tests for LST-1 and MAGIC:

\begin{enumerate}
  \item[(a)] Intrinsic–model test (LST-1).  
    We compare three intrinsic shapes for OP 313: a log‐parabola (LP), 
    a log‐parabola with exponential cutoff (PLEC), and a log‐parabola with cutoff (LPEC).  
    Only the LPEC model yields a physically sensible best–fit for $\alpha\simeq0$.  
    The PLEC curve increases monotonically even at $\alpha<0$, indicating an unphysical preference for ``negative attenuation''.
    We estimated using MC drawn samples the best-fit $p$-values for these different spectral shapes, leaving $\alpha$ free from 0 to 3, obtaining values that range from $p$-value $\sim 0.10$ to $p$-value $\sim 0.15$. 

  \item[(b)] Energy–scale systematics (LST-1).  
    We apply global shifts of $\pm5\%$, $\pm10\%$, and $\pm15\%$ to the IRF energy scale.  
    The most conservative 95\%–CL upper limits on $\alpha$ arise from the \(-15\%\) shift 
    (i.e.\ $0.85\,E_\mathrm{true}$).

  \item[(c)] Intrinsic–model test (MAGIC). 
    As in panel (a), we re‐fit OP~313 data assuming a LP, PLEC, and LPEC.  
    Again, only LPEC gives $\hat\alpha\ge0$, while PLEC suggests $\hat\alpha<0$. 
    We also estimated best-fit $p$-values for the considered spectral shapes, obtaining in this case values between $p$-value $\sim 0.16$ and $p$-value $\sim 0.19$.

  \item[(d)] Energy–scale systematics (MAGIC).
    Similar to panel (b), the $-15\%$ shift yields the most conservative $\alpha_{95}$ limit.
\end{enumerate}

All fits profile over the spectral parameters of OP~313 and the background sources 
in the \textit{Fermi}-LAT ROI for each fixed $\alpha$, treating them as nuisance parameters. In the limit $\alpha \gg 1$, the LP and LPEC attenuation profiles converge, so they yield nearly identical likelihood minima and 95\%–CL upper limits on $\alpha$.

\section{Components of multi-wavelength SED models}
In this appendix, we detail the different components contributing to the broadband emission according to the two scenarios considered for the SED modelling. Figs. \ref{fig:MWL_SED_model_balanced} and \ref{fig:MWL_SED_model_mwlcorr} show the SED models for the spectral-focused and timing-correlation scenarios, respectively.

\subsection{Spectral-focused model}\label{sec:breakdown_model_balanced}
\begin{figure}[h!]
\centering 
\includegraphics[width=\columnwidth]{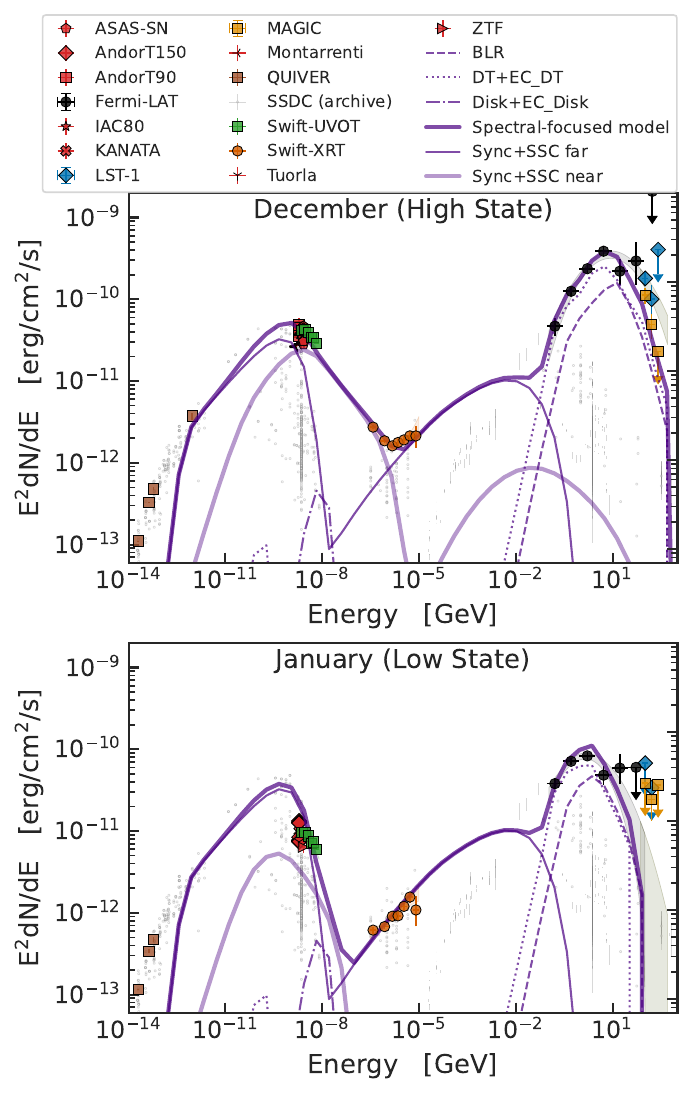}
\caption{\label{fig:MWL_SED_model_balanced}{Spectral-focused} broadband SED models of OP~313. \textit{Top}: {Model} during the December 2023 flare. \textit{Bottom}: {Model} during the January 2024 low state. Different markers represent data at different wavelengths as indicated in the legend. Lines of different styles correspond to the different contributions to the multi-wavelength emission as detailed in the legend. Grey markers correspond to archival data extracted from the SSDC database (\url{https://tools.ssdc.asi.it/SED/}). All flux points have been corrected for extinction, hydrogen absorption and EBL attenuation. }
\end{figure}

The first scenario aims to accurately reproduce the broadband emission, as presented in Figure~\ref{fig:MWL_SED_model_balanced}, where the different contributions to the emission are represented in detail. The $\gamma$-ray emission is explained as EC radiation from the near region, mainly from the DT and the BLR both during the flare in December 2023 and in January 2024, when the source was in low state. The inner region is similarly responsible for an enhanced UV and X-ray emission (particularly below $\sim 1\,$ keV during the flare). The far region on the other side is responsible for most of the radio-to-optical emission in both cases, as well as a hard-spectrum X-ray component (mainly synchrotron-self-Compton) that completely dominates the emission in that band during the low state. The direct emission of the DT and BLR at IR and optical wavelengths is totally sub-dominant with respect to the bright synchrotron emission and therefore, negligible.

\subsection{Timing-correlation model}\label{sec:breakdown_model_mwl_correlations}
\begin{figure}[h!]
\centering 
\includegraphics[width=\columnwidth]{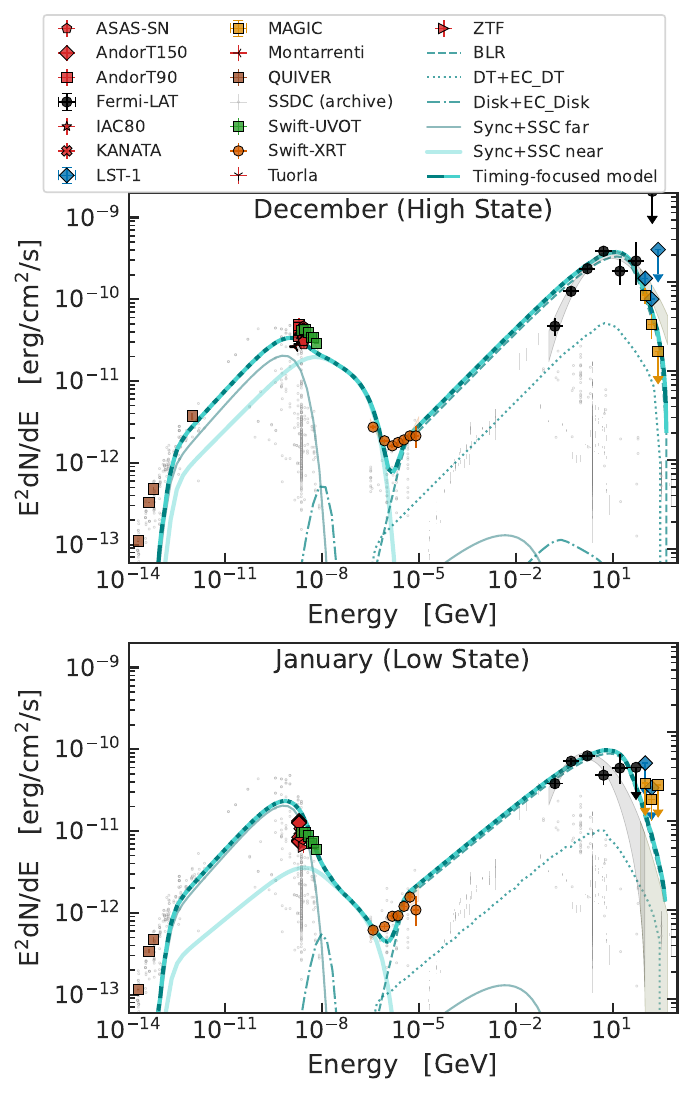}
\caption{\label{fig:MWL_SED_model_mwlcorr}{Timing-correlation broadband} SED models of OP~313. \textit{Top}: {Model} during the December 2023 flare. \textit{Bottom}: {Model} during the January 2024 low state. Same description as Fig.~\ref{fig:MWL_SED_model_balanced}.}

\end{figure}

\begin{figure}[h!]
\centering 
\includegraphics[width=0.9\linewidth]{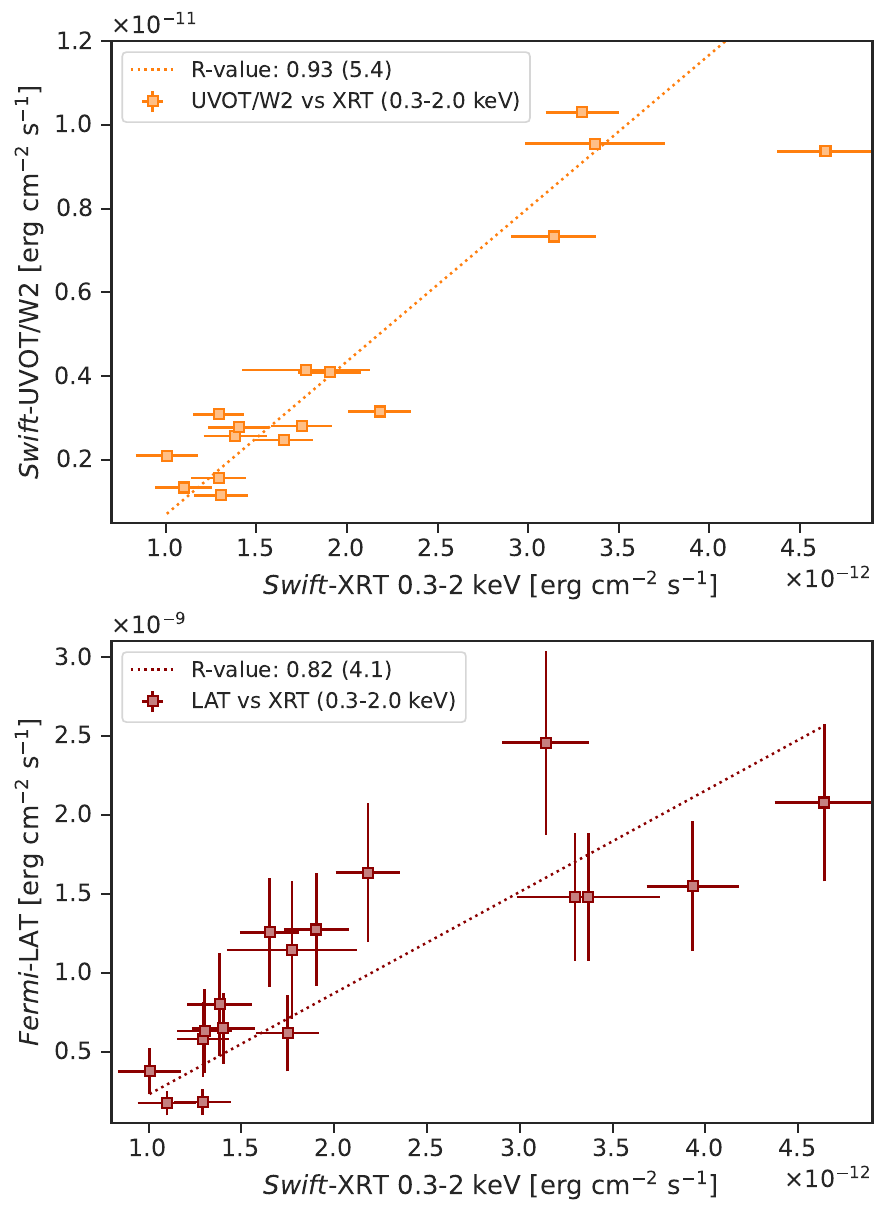}
\caption{\label{fig:uv_xray_corr} Flux-flux correlations between the \textit{Swift}-UVOT/W2 and \textit{Swift}-XRT 0.3-2\,keV fluxes (\textit{left}) and between the \textit{Fermi}-LAT and \textit{Swift}-XRT 0.3-2\,keV fluxes (\textit{right}). The correlation is quantified using the Pearson R coefficient whose value is shown in the respective panels. For the correlation between the \textit{Swift}-UVOT/W2 and \textit{Swift}-XRT 0.3-2\,keV fluxes measured within $\pm2\,h$ are considered, while for the correlation between the \textit{Fermi}-LAT and \textit{Swift}-XRT 0.3-2\,keV fluxes we correlate measurements within a time window of 1\,day (that is the binning of the \textit{Fermi}-LAT light curve).}
\end{figure}

This scenario is motivated by the tight correlations shown in Fig.~\ref{fig:uv_xray_corr}. The X-ray and $\gamma$-ray emission is coming from the near, compact region via SSC and EC (and some synchrotron contribution at soft X-rays during the flare in December). The optical-IR emission is also dominated by this region in December, with a mild contribution at the lowest energies from the far region, that becomes dominant in January 2024. This far region is the one responsible for the much less variable radio emission in both periods. While it is totally consistent with the time correlations observed, it fails to reproduce the SED slope at a few MeVs. The BLR and DT are completely outshined by the synchrotron emission in the optical and IR, respectively. Figure~\ref{fig:MWL_SED_model_mwlcorr} shows in detail the exact contribution of each component of the blazar.

\onecolumn

\begin{acknowledgements}
 We gratefully acknowledge financial support from the following agencies and organisations:
Conselho Nacional de Desenvolvimento Cient\'{\i}fico e Tecnol\'{o}gico (CNPq) Grant 309053/2022-6 and Funda\c{c}\~{a}o de Amparo \`{a} Pesquisa do Estado do Rio de Janeiro (FAPERJ) Grants E-26/200.532/2023 and E-26/211.342/2021, Funda\c{c}\~{a}o de Amparo \`{a} Pesquisa do Estado de S\~{a}o Paulo (FAPESP), Funda\c{c}\~{a}o de Apoio \`{a} Ci\^encia, Tecnologia e Inova\c{c}\~{a}o do Paran\'a - Funda\c{c}\~{a}o Arauc\'aria, Ministry of Science, Technology, Innovations and Communications (MCTIC), Brasil;
Ministry of Education and Science, National RI Roadmap Project DO1-153/28.08.2018, Bulgaria;
Croatian Science Foundation (HrZZ) Project IP-2022-10-4595, Rudjer Boskovic Institute, University of Osijek, University of Rijeka, University of Split, Faculty of Electrical Engineering, Mechanical Engineering and Naval Architecture, University of Zagreb, Faculty of Electrical Engineering and Computing, Croatia;
Ministry of Education, Youth and Sports, MEYS  LM2023047, EU/MEYS CZ.02.1.01/0.0/0.0/16\_013/0001403, CZ.02.1.01/0.0/0.0/18\_046/0016007, CZ.02.1.01/0.0/0.0/16\_019/0000754, CZ.02.01.01/00/22\_008/0004632 and CZ.02.01.01/00/23\_015/0008197 Czech Republic;
CNRS-IN2P3, the French Programme d’investissements d’avenir and the Enigmass Labex, 
This work has been done thanks to the facilities offered by the Univ. Savoie Mont Blanc - CNRS/IN2P3 MUST computing center, France;
Max Planck Society, German Bundesministerium f{\"u}r Forschung, Technologie und Raumfahrt (Verbundforschung / ErUM), Deutsche Forschungsgemeinschaft (SFBs 876 and 1491), Germany;
Istituto Nazionale di Astrofisica (INAF), Istituto Nazionale di Fisica Nucleare (INFN), Italian Ministry for University and Research (MUR), and the financial support from the European Union -- Next Generation EU under the project IR0000012 - CTA+ (CUP C53C22000430006), announcement N.3264 on 28/12/2021: ``Rafforzamento e creazione di IR nell’ambito del Piano Nazionale di Ripresa e Resilienza (PNRR)'';
ICRR, University of Tokyo, JSPS, MEXT, Japan;
JST SPRING - JPMJSP2108;
Narodowe Centrum Nauki, grant number 2023/50/A/ST9/00254, Poland;
The Spanish groups acknowledge the Spanish Ministry of Science and Innovation and the Spanish Research State Agency (AEI) through the government budget lines
PGE2022/28.06.000X.711.04,
28.06.000X.411.01 and 28.06.000X.711.04 of PGE 2023, 2024 and 2025,
and grants PID2019-104114RB-C31,  PID2019-107847RB-C44, PID2019-104114RB-C32, PID2019-105510GB-C31, PID2019-104114RB-C33, PID2019-107847RB-C43, PID2019-107847RB-C42, PID2019-107988GB-C22, PID2021-124581OB-I00, PID2021-125331NB-I00, PID2022-136828NB-C41, PID2022-137810NB-C22, PID2022-138172NB-C41, PID2022-138172NB-C42, PID2022-138172NB-C43, PID2022-139117NB-C41, PID2022-139117NB-C42, PID2022-139117NB-C43, PID2022-139117NB-C44, PID2022-136828NB-C42, PDC2023-145839-I00 funded by the Spanish MCIN/AEI/10.13039/501100011033 and by ERDF/EU and NextGenerationEU PRTR; the "Centro de Excelencia Severo Ochoa" program through grants no. CEX2020-001007-S, CEX2021-001131-S; the "Unidad de Excelencia Mar\'ia de Maeztu" program through grants no. CEX2019-000918-M, CEX2020-001058-M; the "Ram\'on y Cajal" program through grants RYC2021-032991-I  funded by MICIN/AEI/10.13039/501100011033 and the European Union “NextGenerationEU”/PRTR and RYC2020-028639-I; the "Juan de la Cierva-Incorporaci\'on" program through grant no. IJC2019-040315-I and "Juan de la Cierva-formaci\'on"' through grant JDC2022-049705-I; and the support and funding provided by the Viera y Clavijo 2025 Fellowship Program from Gobierno de Canarias and Universidad de La Laguna. They also acknowledge the "Atracci\'on de Talento" program of Comunidad de Madrid through grant no. 2019-T2/TIC-12900; the project "Tecnolog\'ias avanzadas para la exploraci\'on del universo y sus componentes" (PR47/21 TAU), funded by Comunidad de Madrid, by the Recovery, Transformation and Resilience Plan from the Spanish State, and by NextGenerationEU from the European Union through the Recovery and Resilience Facility; “MAD4SPACE: Desarrollo de tecnolog\'ias habilitadoras para estudios del espacio en la Comunidad de Madrid" (TEC-2024/TEC-182) project funded by Comunidad de Madrid; the La Caixa Banking Foundation, grant no. LCF/BQ/PI21/11830030; Junta de Andaluc\'ia under Plan Complementario de I+D+I (Ref. AST22\_0001) and Plan Andaluz de Investigaci\'on, Desarrollo e Innovaci\'on as research group FQM-322; Project ref. AST22\_00001\_9 with funding from NextGenerationEU funds; the “Ministerio de Ciencia, Innovaci\'on y Universidades”  and its “Plan de Recuperaci\'on, Transformaci\'on y Resiliencia”; “Consejer\'ia de Universidad, Investigaci\'on e Innovaci\'on” of the regional government of Andaluc\'ia and “Consejo Superior de Investigaciones Cient\'ificas”, Grant CNS2023-144504 funded by MICIU/AEI/10.13039/501100011033 and by the European Union NextGenerationEU/PRTR,  the European Union's Recovery and Resilience Facility-Next Generation, in the framework of the General Invitation of the Spanish Government's public business entity Red.es to participate in talent attraction and retention programmes within Investment 4 of Component 19 of the Recovery, Transformation and Resilience Plan; Junta de Andaluc\'{\i}a under Plan Complementario de I+D+I (Ref. AST22\_00001), Plan Andaluz de Investigaci\'on, Desarrollo e Innovación (Ref. FQM-322). ``Programa Operativo de Crecimiento Inteligente" FEDER 2014-2020 (Ref.~ESFRI-2017-IAC-12), Ministerio de Ciencia e Innovaci\'on, 15\% co-financed by Consejer\'ia de Econom\'ia, Industria, Comercio y Conocimiento del Gobierno de Canarias; the "CERCA" program and the grants 2021SGR00426 and 2021SGR00679, all funded by the Generalitat de Catalunya; and the European Union's NextGenerationEU (PRTR-C17.I1). This research used the computing and storage resources provided by the Port d'Informaci\'o Cient\'ifica (PIC) data center.
State Secretariat for Education, Research and Innovation (SERI) and Swiss National Science Foundation (SNSF), Switzerland;
The research leading to these results has received funding from the European Union's Seventh Framework Programme (FP7/2007-2013) under grant agreements No~262053 and No~317446;
This project is receiving funding from the European Union's Horizon 2020 research and innovation programs under agreement No~676134;
ESCAPE - The European Science Cluster of Astronomy \& Particle Physics ESFRI Research Infrastructures has received funding from the European Union’s Horizon 2020 research and innovation programme under Grant Agreement no. 824064.

We would like to thank the Instituto de Astrof\'{\i}sica de Canarias for the excellent working conditions at the Observatorio del Roque de los Muchachos in La Palma. The financial support of the German BMFTR, MPG and HGF; the Italian INFN and INAF; the Swiss National Fund SNF; the grants PID2022-136828NB-C41, PID2022-137810NB-C22, PID2022-138172NB-C41, PID2022-138172NB-C42, PID2022-138172NB-C43, PID2022-139117NB-C41, PID2022-139117NB-C42, PID2022-139117NB-C43, PID2022-139117NB-C44, CNS2023-144504 funded by the Spanish MCIN/AEI/ 10.13039/501100011033 and "ERDF A way of making Europe; the Indian Department of Atomic Energy; the Japanese ICRR, the University of Tokyo, JSPS, and MEXT; the Bulgarian Ministry of Education and Science, National RI Roadmap Project DO1-400/18.12.2020 and the Academy of Finland grant nr. 320045 is gratefully acknowledged. This work has also been supported by Centros de Excelencia ``Severo Ochoa'' y Unidades ``Mar\'{\i}a de Maeztu'' program of the Spanish MCIN/AEI/ 10.13039/501100011033 (CEX2019-000918-M, CEX2021-001131-S, CEX2024001442-S), by AST22\_00001\_9 with funding from NextGenerationEU funds and by the CERCA institution and grants 2021SGR00426, 2021SGR00607 and 2021SGR00773 of the Generalitat de Catalunya; by the Croatian Science Foundation (HrZZ) Project IP-2022-10-4595 and the University of Rijeka Project uniri-prirod-18-48; by the Deutsche Forschungsgemeinschaft (SFB1491) and by the Lamarr-Institute for Machine Learning and Artificial Intelligence; by the Polish Ministry Of Education and Science grant No. 2021/WK/08; by the European Union (ERC, MicroStars, 101076533); and by the Brazilian MCTIC, the CNPq Productivity Grant 309053/2022-6 and FAPERJ Grants E-26/200.532/2023 and E-26/211.342/2021.

Some of the data are based on observations collected at the Observatorio de Sierra Nevada; which is owned and operated by the Instituto de Astrof\'isica de Andaluc\'ia (IAA-CSIC).
This article is based on observations made with the IAC80 operated on the island of Tenerife by the Instituto de Astrofísica de Canarias in the Spanish Observatorio del Teide.
This work makes use of observations from the Las Cumbres Observatory global telescope network.
This work makes use of data from the All-Sky Automated Survey for Supernovae (ASAS-SN).
This work has made use of data from the Joan Oró Telescope (TJO) of the Montsec Observatory (OdM), which is owned by the Catalan Government and operated by the Institute for Space Studies of Catalonia (IEEC).
Based on observations obtained with the Samuel Oschin Telescope 48-inch and the 60-inch Telescope at the Palomar Observatory as part of the Zwicky Transient Facility project. ZTF is supported by the National Science Foundation under Grants No. AST-1440341 and AST-2034437 and a collaboration including current partners Caltech, IPAC, the Oskar Klein Center at Stockholm University, the University of Maryland, University of California, Berkeley , the University of Wisconsin at Milwaukee, University of Warwick, Ruhr University, Cornell University, Northwestern University and Drexel University. Operations are conducted by COO, IPAC, and UW.
The Submillimeter Array is a joint project between the Smithsonian Astrophysical Observatory and the Academia Sinica Institute of Astronomy and Astrophysics and is funded by the Smithsonian Institution and the Academia Sinica. Maunakea, the location of the SMA, is a culturally important site for the indigenous Hawaiian people; we are privileged to study the cosmos from its summit.
Partly based on observations with the 100-m telescope of the MPIfR (Max- Planck-Institut für Radioastronomie) at Effelsberg. Observations with the 100-m radio telescope at Effelsberg have received funding from the European Union’s Horizon 2020 research and innovation programme under grant agreement No 101004719 (ORP).

M.N.R. acknowledges support from the Agencia Estatal de Investigación del Ministerio de Ciencia, Innovación y Universidades (MCIU/AEI) under grant PARTICIPACIÓN DEL IAC EN EL EXPERIMENTO AMS and the European Regional Development Fund (ERDF) with reference PID2022-137810NB-C22/DIO 10.13039/501100011033.

J.B. thanks the supports from JSPS KAKENHI Grant Number 1268737.

A.A.-E. acknowledges support from the Deutsche Forschungs gemeinschaft (DFG, German Research Foundation) under Germany’s Excellence Strategy – EXC-2094 – 390783311.

This work benefited from the support of the project COCOA-NuGETs ANR-23-CE31-0026 of the French National Research Agency (ANR).  \\

\noindent \textit{Author contributions:} A. Arbet-Engels: MAGIC data taking coordination, data analysis and systematic calculation, modelling, paper edition. J. Baxter: work on optical observation coordination and EBL measurement, paper edition. A. Briscioli: MAGIC analysis cross-check. L. Heckmann: MAGIC data analysis and systematic calculation, paper edition. D. Morcuende: LST-1 data analysis and systematics calculation, paper edition. M. Nievas Rosillo: \textit{Swift} observation coordination, multi-wavelength data analysis and modelling, systematics calculation, EBL measurements and modeling, paper edition. J.~Otero-Santos: Optical and radio observation coordination, LST-1 data analysis, multi-wavelength data analysis and modeling, paper edition. D. A. Sanchez: project coordination, work on EBL measurement, paper edition.
The rest of the authors have contributed in one or several of the following ways: design, construction, maintenance and operation of the instrument(s) used to acquire the data; preparation and/or evaluation of the observation proposals; data acquisition, processing, calibration and/or reduction; production of analysis tools and/or related Monte Carlo simulations; discussion and approval of the contents of the draft.
\end{acknowledgements}

\end{appendix}

\end{document}